\documentclass[11pt,a4paper]{article}
\usepackage{jheppub}
\usepackage{amsmath}
\usepackage{amssymb}

\newcommand{\fldZ}{\mathcal{Z}}

\newcommand{\fldD}{\mathcal{D}}

\newcommand{\transcendentalitylevel}{transcendentality}
\newcommand{\transcendentalitylevels}{transcendentalities}

\newcommand{\alg}[1]{\mathfrak{#1}}

\newcommand{\gym}{g\indups{YM}}
\newcommand{\indups}[1]{_{\mathrm{\scriptscriptstyle #1}}}
\newcommand{\gaba}{\gamma\inddowns{ABA}}
\newcommand{\inddowns}[1]{^{\mathrm{\scriptscriptstyle #1}}}

\newcommand{\womega}{\omega}

\newcommand{\ABA}{\mbox{\tiny ABA}}
\newcommand{\Wrap}{\mathrm{wrap}}

\newcommand{\Tr}{{\rm Tr \,}}
\newcommand{\Op}{\mathcal{O}}

\newcommand{\cN}{\mathcal{N}}

\def\HBS{\mathbb S}
\def\HS{S}
\def\reciP{\mathcal{P}}
\def\reciT{\mathcal{T}}
\def\cP{\mathcal{P}}
\def\cT{\mathcal{T}}

\newcommand{\beq}{\begin{equation}}
\newcommand{\eeq}{\end{equation}}
\newcommand{\beqa}{\begin{eqnarray}}
\newcommand{\eeqa}{\end{eqnarray}}

\newcommand{\h}{{\mathrm{h}}}

\newcommand{\sfrac}[2]{{\textstyle\frac{#1}{#2}}}

\newcommand{\M}{M}

\def\z#1{{{\zeta_#1}}}
\def\zp#1#2{{{\zeta_#1^#2}}}
\def\zp#1#2{{{\zeta_#1^#2}}}

\title{Six-loop anomalous dimension of twist-two operators in planar $\cN=4$ SYM theory}

\author[a]{Christian Marboe}
\author[b,c]{~~~Vitaly Velizhanin}
\author[a]{~~~Dmytro Volin}

\affiliation[a]{School of Mathematics, Trinity College Dublin, College
  Green, Dublin 2, Ireland.}

\affiliation[b]{
Institut f{\"u}r  Mathematik und Institut f{\"u}r Physik,
Humboldt-Universit{\"a}t zu Berlin,
IRIS Adlershof, Zum Gro\ss{}en Windkanal 6,
12489 Berlin, Germany.
}

\affiliation[c]{
Theoretical Physics Division,
NRC "Kurchatov Institute",
Petersburg Nuclear Physics Institute,
Orlova Roscha, Gatchina,
188300 St.~Petersburg, Russia.}

\emailAdd{marboec@tcd.ie}

\emailAdd{velizh@physik.hu-berlin.de}

\emailAdd{volind@tcd.ie}

\abstract{
We compute the general form of the six-loop anomalous dimension of twist-two operators with arbitrary spin in planar $\cN=4$ SYM theory. First we find the contribution from the asymptotic Bethe ansatz. Then we reconstruct the wrapping terms from the first 35 even spin values of the full six-loop anomalous dimension computed using the quantum spectral curve approach. The obtained anomalous dimension satisfies all known constraints coming from the BFKL equation, the generalised double-logarithmic equation, and the small spin expansion.
}

\preprint{
          {\tiny{HU-Mathematik-P-2014-37}} \\[-.95ex]
          {\hspace*{135.6mm}\tiny{HU-EP-14/60}} 
          }

\arxivnumber{1412.4762}

\begin{document}

\maketitle

\section{Introduction}
\label{sec:intro}

The anomalous dimension of composite gauge-invariant operators in $\cN=4$ SYM theory can be calculated with the help of integrability. Integrability in the context of AdS/CFT-correspondence~\cite{Maldacena:1997re,Gubser:1998bc,Witten:1998qj} was found from the study of the single-trace operators~\cite{Berenstein:2002jq} in the leading order of perturbative theory in ref.~\cite{Minahan:2002ve}\footnote{Earlier, similar integrability was discovered in quantum chromodynamics in the Regge limit~\cite{Lipatov:1993yb,Lipatov:1994xy,Faddeev:1994zg} and for some operators~\cite{Braun:1998id}.}.
Generalisation to higher orders together with the studies of integrable structures from the superstring theory side, started in ref.~\cite{Bena:2003wd}, allowed formulating all-loop asymptotic Bethe equations
\cite{Beisert:2003tq,Beisert:2003yb,Beisert:2003jj,Beisert:2003ys,Serban:2004jf,Kazakov:2004qf,Beisert:2004hm,Arutyunov:2004vx,Staudacher:2004tk,
Beisert:2005fw,Beisert:2005tm,Janik:2006dc,Hernandez:2006tk,Arutyunov:2006iu,Beisert:2006ib,Eden:2006rx,Bern:2006ew,Beisert:2006ez}.
For the operators of finite length the asymptotic Bethe equations give a non-complete result due to the appearance of wrapping effects~\cite{SchaferNameki:2006ey,Kotikov:2007cy}.
Again, the computations of wrapping corrections can be performed using integrability~\cite{Bajnok:2008bm,Bajnok:2008qj,Bajnok:2009vm,Lukowski:2009ce,Gromov:2009tv,Gromov:2009bc,Gromov:2009zb,
Bombardelli:2009ns,Arutyunov:2009ax,Arutyunov:2009ur,Arutyunov:2010gb,Balog:2010xa,Balog:2010vf,Bajnok:2010ud}.
Independent tests of the obtained results were performed with the perturbative field theory computations in refs.~\cite{Fiamberti:2007rj,Fiamberti:2008sh,Velizhanin:2008jd,Velizhanin:2008pc}.
Moreover, the results for twist-two and spin $\M$ operators passed very non-trivial tests coming from the Balitsky-Fadin-Kuraev-Lipatov (BFKL)~\cite{Lipatov:1976zz,Kuraev:1977fs,Balitsky:1978ic}
and generalised double-logarithmic~\cite{Gorshkov:1966ht,Gorshkov:1966hu,Gorshkov:1966qd,Kirschner:1982qf,Kirschner:1982xw,Kirschner:1983di,Velizhanin:2011pb} equations which impose all-loop constrains on the structure of the result analytically continued to negative values of $\M$. Recently, the leading pole BFKL constraint was shown~\cite{Alfimov:2014bwa} to analytically follow from the quantum spectral curve (QSC) equations~\cite{Gromov:2013pga,Gromov:2014caa} which is the most conscise currently known way to describe the spectrum of anomalous dimensions based on integrability.

In principle, we can do the inverse: we can use constraints from the BFKL and the generalised double-logarithmic equations to reconstruct the wrapping corrections if the part of the full anomalous dimension related to the asymptotic Bethe equations is already known. This was done at five loops by one of the authors in ref.~\cite{Velizhanin:2013vla}. However, one could not obtain any reasonable results at six loops in this way.

Fortunately, an  efficient method for computing the anomalous dimension of operators with arbitrary but fixed twist and spin was developed recently by two of the authors~\cite{Marboe:2014gma}. This method performs a perturbative solution of the QSC equations and it can, in principle, produce results to any loop order. In this paper we combine results of~\cite{Marboe:2014gma} together with previous findings to reconstruct a general form of the six-loop anomalous dimension for twist-two operators with arbitrary spin $M$ and use all available constraints to check its correctness.

In section~\ref{sec:ABA} we compute the part of the six-loop anomalous dimension for twist-two operators coming from the ABA.
In section~\ref{sec:finiteS} we briefly describe the method of computation of the six-loop anomalous dimension for twist-two operators with fixed spin.
In section~\ref{sec:L6} we reconstruct the general form of the six-loop anomalous dimension from the first 35 even values, obtained with the method described in section~\ref{sec:finiteS}.
In section~\ref{sec:weak} we provide the constraints which will be used to verify the obtained result, together with the description of their origin.
In appendices we give the results for the most complicated parts of the six-loop anomalous dimension and its analytic continuation at $\M=-2+\omega$.

\section{The six-loop anomalous dimension from Bethe ansatz}\label{sec:ABA}

Twist-two operators belong to the $\mathfrak{sl}(2)$ sub-sector of the theory. In this sector, the highest-weight states consist of two scalar fields $\fldZ$ and $\M$ \footnote{$\M$ is the value of the Lorentz spin. Hence the notation $S$ is often used instead of $M$. We choose to use~$M$ to not interfere with the notation for harmonic sums.} covariant derivatives~$\fldD$
\begin{equation}
\label{twisttwo}
\Tr \left( \fldZ\, \fldD^\M\, \fldZ\,\right)\,,
\end{equation}
where $M$ should be even. In the spin chain picture which is valid at one loop at weak coupling, such single-trace operators are identified with the states of the non-compact $\alg{sl}(2)$
spin $=-\sfrac{1}{2}$ length-two Heisenberg magnet with $\M$ excitations.

We denote the total scaling dimension of these states as%
\begin{equation}
\label{dimension}
\Delta=2+\M+\gamma(g)\, ,
\qquad {\rm with} \qquad
\gamma(g)=\sum_{\ell=1}^\infty  \gamma^{}_{2\ell}\,g^{2\ell}\, .
\end{equation}
Here, $\gamma(g)$ is the anomalous part of the dimension
depending on the coupling constant
\begin{equation}
\label{convention}
g^2=\frac{\lambda}{16\,\pi^2}\, ,
\end{equation}
and $\lambda=N\, \gym^2$ is the 't Hooft coupling constant. The anomalous dimension $\gamma(g)$ may be determined up to the three-loop order $\Op(g^6)$ with help of the asymptotic Bethe ansatz \cite{Staudacher:2004tk}.

The asymptotic Bethe equations for the $\mathfrak{sl}(2)$ operators can be found in \cite{Beisert:2005fw,Beisert:2006ez}
\begin{equation}
\label{sl2eq}
\left(\frac{x^+_k}{x^-_k}\right)^L=\prod_{\substack{j=1\\j \neq k}}^\M
\frac{x_k^--x_j^+}{x_k^+-x_j^-}\,
\frac{1-g^2/x_k^+x_j^-}{1-g^2/x_k^-x_j^+}\,
\exp\left(2\,i\,\theta(u_k,u_j)\right),
\qquad
\prod_{k=1}^M \frac{x^+_k}{x^-_k}=1\, ,
\end{equation}
where the variables $x^{\pm}_k$ are related to $u_k$ through
\begin{equation}\label{definition x}
x_k^{\pm}=x(u_k^\pm)\, ,
\qquad
u^\pm=u\pm\tfrac{i}{2}\, ,
\qquad
x(u)=\frac{u}{2}\left(1+\sqrt{1-4\,\frac{g^2}{u^2}}\right).
\end{equation}
The dressing phase $\theta(u,v)$ has been conjectured in~\cite{Beisert:2006ez} and shown to be solution of the crossing equation \cite{Janik:2006dc} in \cite{Arutyunov:2009kf,Volin:2009uv}. To the sixth order in perturbation theory it has the following form
\begin{eqnarray}
\label{4loopphase}
\theta(u_k,u_j) &=&\left(4\,
\z3\,g^6  -40\, \z5 g^8 + 420\, \z7 g^{10} \right) \big(q_2(u_k)\,q_3(u_j)-q_3(u_k)\,q_2(u_j)\big)\nonumber\\[2mm]
&&-\ 8\, \z5 g^{10} \big(q_2(u_k)\,q_5(u_j)-q_5(u_k)\,q_2(u_j)\big)\nonumber\\[2mm]
&&+\ 24\, \z5 g^{10} \big(q_3(u_k)\,q_4(u_j)-q_4(u_k)\,q_3(u_j)\big)
+\Op(g^{12})\, ,\label{DrPhase}
\end{eqnarray}
where $q_r(u)$ are the eigenvalues of the conserved magnon charges,  see \cite{Beisert:2005fw}.
The anomalous dimension is given by
\begin{equation}
\label{dim}
\gaba(g)=2\, g^2\, \sum^{\M}_{k=1}
\left(\frac{i}{x^{+}_k}-\frac{i}{x^{-}_k}\right) =
\sum_{l=1}^\infty g^{2l}\,\gamma^{\ABA}_{2l}(M)
\end{equation}
and we will decompose it at six loops as
\begin{equation}
\gamma_{12}^{\ABA}(M)\ =\
\hat\gamma_{12}\ =\
\hat\gamma_{12}^{\rm rational}
+\hat\gamma_{12}^{\z3}\,\z3
+\hat\gamma_{12}^{\z5}\,\z5
+\hat\gamma_{12}^{\z7}\,\z7\,.
\label{gammaABA}
\end{equation}

At one loop the Bethe roots $u_k$ are given by the zeros of the Hahn polynomial~\cite{Dippel,Eden:2006rx}
\begin{equation}
\label{eq:Wilsonpolynomial}
    P_M(u)={}_3 F_2\left(\left. \begin{array}{c}
    -M, \ M+1,\ \frac{1}{2}+iu \\[3mm]
    1,\  1 \end{array}
    \right| 1\right) \,.
\end{equation}

In order to obtain a general expression for the anomalous dimension of twist-two operators for arbitrary $\M$ we solve eq. (\ref{sl2eq}) perturbatively for fixed values of the spin $\M$ and match the coefficients in an appropriate ansatz which assumes the maximal transcendentality principle~\cite{Kotikov:2002ab}\footnote{The hypothesis about the maximal transcendentality principle~\cite{Kotikov:2002ab} was confirmed by direct perturbative calculations at the two-loop order~\cite{Kotikov:2003fb} and then successfully applied at the three-loop order~\cite{Kotikov:2004er}, when corresponding results were obtained in QCD~\cite{Moch:2004pa,Vogt:2004mw}.}.
The basis for the ansatz is formed from the harmonic sums defined by the following recurrent procedure (see \cite{Vermaseren:1998uu})
\begin{eqnarray} \label{vhs1}
S_a (M)&=&\sum^{M}_{j=1} \frac{(\mbox{sgn}(a))^{j}}{j^{\vert a\vert}}\, ,
\\[3mm]
S_{a_1,\ldots,a_n}(M)&=&\sum^{M}_{j=1} \frac{(\mbox{sgn}(a_1))^{j}}{j^{\vert a_1\vert}}
\,S_{a_2,\ldots,a_n}(j)\, .\label{vhs}
\end{eqnarray}
The \transcendentalitylevel{} $k$ of each sum $S_{a_1,\ldots,a_n}$ is given by the sum of the absolute values of its indices
\beq
k=\vert a_1 \vert +\ldots \vert a_n \vert\,,
\eeq
and the \transcendentalitylevel{} of a product of harmonic sums is equal to the sum of the \transcendentalitylevels{} of its constituents.
According to the maximal transcendentality principle~\cite{Kotikov:2002ab,Kotikov:2003fb,Kotikov:2004er,Kotikov:2007cy} the anomalous dimension of twist-two operators can contain only harmonic sums with maximal transcendentality at a given order of perturbative theory. The number of harmonic sums entering into this basis at the $\ell$-loop order for the \transcendentalitylevel{} $k=2\ell-1$ is equal to $((1 - \sqrt{2})^k + (1 + \sqrt{2})^k)/2$, and at six loops we will need more than 8800 sums, see Table~\ref{table:nums}.

\begin{table}[t]
\centering
\begin{tabular}{|c|c|c|c|c|c|c|c|c|c|c}
  \hline
                             &                   &       &       &         &       &          &         &       &      \\[-2mm]
         Contribution        & Rational          & $\z3$ & $\z5$ & $\zp{3}{2}$ & $\z7$ & $\z5\z3$ & $\zp{3}{3}$ & $\z9$ & Total\\[2mm]
  \hline
                             &                   &       &       &         &       &          &         &       &      \\[-2mm]
  Transcendentality of basis & 11                &   8   & 6     & 5       & 4     & 3        & 2       & 2     &\\[2mm]
  \hline
                             &                   &       &       &         &       &          &         &       &      \\[-2mm]
  $\gamma_{\ABA}$        &  8119             & 577   & 99    &         & 17    &          &         &       & 8812\\[1mm]
  $\reciP_{\ABA}$        &  1024             & 128   & 32    &         & 8     &          &         &       & 1192\\[2mm]
  $\gamma_{\rm{Wrapping}}$   &  1393             & 99    & 17    & 7       & 3     & 1        & 1       & 1     & 1522\\[1mm]
  $\reciP_{\rm{Wrapping}}$   &  256              & 32    & 8     & 4       & 2     & 1        & 1       & 1     &  305\\[2mm]
  \hline
\end{tabular}
\caption{The number of harmonic sums in the basis for different contributions.}
\label{table:nums}
\end{table}

Fortunately, the generalised Gribov-Lipatov reciprocity~\cite{Dokshitzer:2005bf,Dokshitzer:2006nm} enters the game and allows us to significantly reduce the dimension of the basis.  Define the reciprocity-respecting function $\reciP^{\mbox{\tiny ABA}}(\M)$~\cite{Dokshitzer:2005bf,Dokshitzer:2006nm,Basso:2006nk} by the relation
\begin{equation} \label{Pfunction}
\gaba(M) = \reciP^{\ABA} \left(M+\frac{1}{2} \gaba(M) \right)\,.
\end{equation}
The reciprocity-respecting splitting function ${\mathcal P}(x)$~\cite{Dokshitzer:2005bf,Dokshitzer:2006nm}, related to $\mathcal P(M)$ through a Mellin transformation, should satisfy
the Gribov-Lipatov relation~\cite{Gribov:1972ri}
\begin{equation}\label{GrLipRel}
{\mathcal P}(x)=-\,x\,{\mathcal P}\!\left(\frac{1}{x}\right)\,
\end{equation}
at all orders of perturbation theory.

The practical output which we will use is that the reciprocity function $\reciP^{\mbox{\tiny ABA}}(\M)$ should be  expressed only in terms of the binomial sums (see \cite{Vermaseren:1998uu})
\beq\label{BinomialSums}
\HBS_{i_1,\ldots,i_k}(N)=(-1)^N\sum_{j=1}^{N}(-1)^j\binom{N}{j}\binom{N+j}{j}\HS_{i_1,...,i_k}(j)\,,
\eeq
which is the necessary and sufficient condition to satisfy \eqref{GrLipRel} \cite{Lukowski:2009ce}\footnote{Historically, this statement was understood using a basis different from, but equivalent to, the basis of binomial sums \cite{Dokshitzer:2006nm,Beccaria:2007bb,Beccaria:2009vt,Beccaria:2009eq}. Relations between the binomial and the nested harmonic sums can be found in the ancillary files of the arXiv version this paper or on the web-page \href{http://thd.pnpi.spb.ru/~velizh/6loop/}{\texttt{http://thd.pnpi.spb.ru/\textasciitilde velizh/6loop/}}.}.

The binomial sums come both from the perturbative field theory calculations of the anomalous dimensions of twist-two operators~\cite{Moch:2004pa,Vogt:2004mw} and from the solution of the Baxter equation for the corresponding spin chain~\cite{Dippel,Kotikov:2008pv}. One of the interesting features of these sums is that they are defined only for \textit{positive} values of the indices $i_1,\ldots,i_k$.
The number of such sums, which will form the basis, for \transcendentalitylevel{} $k$ is equal to $2^{k-1}$ and at six loops we will have about 1000 binomial harmonic sums (see Table~\ref{table:nums}). Thus, the basis is significantly reduced and, though still huge, can be managed.

In this paper, we first find the result in terms of  $\reciP(\M)$ and then reconstruct the anomalous dimension from \eqref{Pfunction}: upon substituting the perturbative expansion \eqref{dimension}, one finds
\begin{equation}
\reciP^{\ABA}(M)=\sum_{l=1}^\infty g^{2l}\,\reciP^{\ABA}_{2l}(M)\,.\label{PABA}
\end{equation}

From eqs. (\ref{Pfunction}) and~(\ref{PABA}) one can find that at the six-loop order the reciprocity function $\reciP_{12}$ is related to the anomalous dimension $\gamma_{12}$ (see Appendix B of ref.~\cite{Beccaria:2009eq} for five loops):
\beqa
\reciP_{12}^{\ABA}(\M)&=&\hat\reciP_{12}\ =\
\hat\reciP_{12}^{\rm rational}
+\hat\reciP_{12}^{\z3}\z3
+\hat\reciP_{12}^{\z5}\z5
+\hat\reciP_{12}^{\z7}\z7\,,\label{PABAExp}\\[2mm]
\hat\reciP_{12}^{\rm rational}&=&
\hat\gamma_{12}^{\rm rational}
-\frac{1}{4} \left(\hat\gamma_6^2+2\, \hat\gamma_4 \hat\gamma_8^{\rm rational}
+2\, \hat\gamma_2 \hat\gamma_{10}^{\rm rational}\right)'\nonumber\\
&&+\frac{1}{24} \left(\hat\gamma_4^3+6\, \hat\gamma_2 \hat\gamma_6 \hat\gamma_4+3\, \hat\gamma_2^2 \hat\gamma_8^{\rm rational}\right)''\nonumber\\
&&-\frac{1}{96} \hat\gamma_2^2 \left(3\, \hat\gamma_4^2+2\, \hat\gamma_2 \hat\gamma_6\right)'''
+\frac{1}{384}\left(\hat\gamma_2^4 \hat\gamma_4\right)''''
-\frac{1}{23040}\left(\hat\gamma _2^6\right)'''''\,,
\label{P12}\\
\hat\reciP_{12}^{\z3}&=&
\hat\gamma_{12}^{\z3}
-\frac{1}{2} \left(\hat\gamma_4 \hat\gamma_8^{\z3}+\hat\gamma_2 \hat\gamma_{10}^{\z3}\right)'
+\frac{1}{8} \left(\hat\gamma_2^2 \hat\gamma_8^{\z3}\right)''\,,
\label{Pzt}\\
\hat\reciP_{12}^{\z5}&=&
\hat\gamma_{12}^{\z5}
-\frac{1}{2} \left(\hat\gamma_2 \hat\gamma_{10}^{\z5}\right)'\,,
\label{Pzf}\\
\hat\reciP_{12}^{\z7}&=&
\hat\gamma_{12}^{\z7}\,,
\label{Pzs}
\eeqa
where each prime marks a derivative with respect to $\M$.

The rational and $\zeta_i$ parts of $\reciP_{12}^{\ABA}(\M)$ are computed separately. To compute the rational part one should fix $2^{11-1}=1024$ coefficients in front of the binomial harmonic sums of transcendentality 11. Hence the same number of solutions of the Bethe equations at fixed $M$ is required. Although we can "switch off" the dressing phase (because we are looking for the rational part of the answer), analytic solution of the Bethe equations for the first 1024 values of $M$ is beyond computer ability. It was possible to solve these equations up to six loops numerically with accuracy of about  $10^{-1000}$. This accuracy is not enough to reconstruct the rational numbers for the anomalous dimension at given $M$, however this is not what we need. One needs to solve equations for the coefficients in front of the harmonic sums. This was done numerically with the help of the \texttt{MATHEMATICA} function \texttt{LinearSolve}. It turned out that the obtained numbers are very close to integers, so their rounding gives the desired result for $\reciP_{12}(M)$, which is listed in Appendix \ref{SpecialSums}.

When Bethe equations include the dressing phase~(\ref{DrPhase}), the first 40 solutions were computed. According to Table~\ref{table:nums}, this is enough to find the general expression for the functions $\hat\reciP^{\z5}_{12}$ and $\hat\reciP^{\z7}_{12}$, which have the following form:
\beqa
\hat\cP_{12}^{\z5}&\!\!=\!&
-64 \Big(
-8\HBS_6
58\HBS_{1,5}
-74\HBS_{2,4}
-8\HBS_{3,3}
-8\HBS_{4,2}
+60\HBS_{5,1}
+40\HBS_{1,1,4}
-78\HBS_{1,2,3}\nonumber\\[2mm]&&\hspace*{-7mm}
-38\HBS_{1,3,2}
+40\HBS_{1,4,1}
-12\HBS_{2,1,3}
+88\HBS_{2,2,2}
-40\HBS_{2,3,1}
-26\HBS_{3,2,1}
+26\HBS_{4,1,1}
-60\HBS_{1,1,2,2}\nonumber\\[2mm]&&\hspace*{-7mm}
+20\HBS_{1,1,3,1}
-20\HBS_{1,2,1,2}
+38\HBS_{1,3,1,1}
-20\HBS_{2,1,1,2}
+16\HBS_{2,1,2,1}
+14\HBS_{2,2,1,1}
-4\HBS_{3,1,1,1}
\Big),\qquad\
\label{ABAz5}\\
\hat\cP_{12}^{\z7}&\!\!=\!&
-3360 \left(
2 \,\HBS_4
+\,\HBS_{1,3}
-\,\HBS_{2,2}
-\,\HBS_{1,2,1}
-\,\HBS_{2,1,1}
\right)
\,.\label{ABAz7}
\eeqa

For the reconstruction of $\hat\reciP^{\z3}_{12}$ we used the \texttt{LLL}-algorithm~\cite{Lenstra:1982}, similar to what was done in the previous works~\cite{Velizhanin:2010cm,Velizhanin:2012nm} by one of the authors (see also ref.~\cite{Velizhanin:2013vla} for detailed explanations).
Its usage is based on the fact that the coefficients in any anomalous dimension of twist-two operators are usually rather simple integers, that is the equation for the coefficients is a linear Diophantine equation.
Moreover, a lot of binomial harmonic sums which belong to the basis of possible terms are absent in the final expression, i.e. their coefficients are zeros.
The \texttt{LLL}-algorithm is realized in many computer algebra systems and for our purposes the \texttt{MATHEMATICA} function \texttt{LatticeReduce} has been used to realize the algorithm.
Firstly, we calculate the values of all 128 terms in the basis with \transcendentalitylevel{} 8 up to $\M=40$.
We hence can write a system of $39$ linear equations for $128$ coefficients which are assumed to be integers.
According to the realization of the \texttt{LLL}-algorithm\footnote{See {\texttt{Application}} in
{\texttt{{http://reference.wolfram.com/mathematica/ref/LatticeReduce.html}}}} we add to the $129\times 129$ unit matrix the transpose matrix, obtained from our system of Diophantine equations (coefficients of $128$ variables plus one free term), and applying the \texttt{LatticeReduce} function of \texttt{MATHEMATICA} to the obtained $(129+39)\times 129$ matrix we get the \texttt{LLL}-reduced matrix, in which we can easily find the result that we are looking for:
\beqa
\hat\cP_{12}^{\z3}&\!\!=\!&
-32 \big(
4\,\HBS_{2,6}
-4\,\HBS_{7,1}
-8\,\HBS_{1,1,6}
+12\,\HBS_{1,2,5}
+3\,\HBS_{1,3,4}
+3\,\HBS_{1,4,3}
+3\,\HBS_{1,5,2}
-13\,\HBS_{1,6,1}\nonumber\\[1.12mm]&&\hspace*{-3mm}
+8\,\HBS_{2,1,5}
-18\,\HBS_{2,2,4}
-7\,\HBS_{2,3,3}
-7\,\HBS_{2,4,2}
+20\,\HBS_{2,5,1}
-11\,\HBS_{3,1,4}
+8\,\HBS_{3,2,3}
-\,\HBS_{3,3,2}\nonumber\\[1.12mm]&&\hspace*{-3mm}
+7\,\HBS_{3,4,1}
-9\,\HBS_{4,1,3}
+7\,\HBS_{4,2,2}
+7\,\HBS_{4,3,1}
-9\,\HBS_{5,1,2}
+21\,\HBS_{5,2,1}
-16\,\HBS_{6,1,1}
-6\,\HBS_{1,1,1,5}\nonumber\\[1.12mm]&&\hspace*{-3mm}
+14\,\HBS_{1,1,2,4}
+8\,\HBS_{1,1,3,3}
+8\,\HBS_{1,1,4,2}
-12\,\HBS_{1,1,5,1}
+5\,\HBS_{1,2,1,4}
-24\,\HBS_{1,2,2,3}
-13\,\HBS_{1,2,3,2}\nonumber\\[1.12mm]&&\hspace*{-3mm}
+19\,\HBS_{1,2,4,1}
-9\,\HBS_{1,3,1,3}
+3\,\HBS_{1,3,2,2}
+9\,\HBS_{1,3,3,1}
-8\,\HBS_{1,4,1,2}
+25\,\HBS_{1,4,2,1}
-19\,\HBS_{1,5,1,1}\nonumber\\[1.12mm]&&\hspace*{-3mm}
+5\,\HBS_{2,1,1,4}
-12\,\HBS_{2,1,2,3}
-7\,\HBS_{2,1,3,2}
+6\,\HBS_{2,1,4,1}
+2\,\HBS_{2,2,1,3}
+26\,\HBS_{2,2,2,2}
-22\,\HBS_{2,2,3,1}\nonumber\\[1.12mm]&&\hspace*{-3mm}
+8\,\HBS_{2,3,1,2}
-33\,\HBS_{2,3,2,1}
+27\,\HBS_{2,4,1,1}
-10\,\HBS_{3,1,1,3}
+10\,\HBS_{3,1,2,2}
+9\,\HBS_{3,2,1,2}
-24\,\HBS_{3,2,2,1}\nonumber\\[1.12mm]&&\hspace*{-3mm}
+10\,\HBS_{3,3,1,1}
-9\,\HBS_{4,1,1,2}
+9\,\HBS_{4,1,2,1}
+23\,\HBS_{4,2,1,1}
-18\,\HBS_{5,1,1,1}
+9\,\HBS_{1,1,1,2,3}
+9\,\HBS_{1,1,1,3,2}\nonumber\\[1.12mm]&&\hspace*{-3mm}
-6\,\HBS_{1,1,1,4,1}
+3\,\HBS_{1,1,2,1,3}
-24\,\HBS_{1,1,2,2,2}
+13\,\HBS_{1,1,2,3,1}
-7\,\HBS_{1,1,3,1,2}
+27\,\HBS_{1,1,3,2,1}\nonumber\\[1.12mm]&&\hspace*{-3mm}
-20\,\HBS_{1,1,4,1,1}
+3\,\HBS_{1,2,1,1,3}
-8\,\HBS_{1,2,1,2,2}
+3\,\HBS_{1,2,1,3,1}
+3\,\HBS_{1,2,2,1,2}
-37\,\HBS_{1,2,2,2,1}\nonumber\\[1.12mm]&&\hspace*{-3mm}
+27\,\HBS_{1,2,3,1,1}
-9\,\HBS_{1,3,1,1,2}
+9\,\HBS_{1,3,1,2,1}
+25\,\HBS_{1,3,2,1,1}
-20\,\HBS_{1,4,1,1,1}
+3\,\HBS_{2,1,1,1,3}\nonumber\\[1.12mm]&&\hspace*{-3mm}
-8\,\HBS_{2,1,1,2,2}
+3\,\HBS_{2,1,1,3,1}
-3\,\HBS_{2,1,2,1,2}
-5\,\HBS_{2,1,2,2,1}
+10\,\HBS_{2,1,3,1,1}
+4\,\HBS_{2,2,1,1,2}\nonumber\\[1.12mm]&&\hspace*{-3mm}
-6\,\HBS_{2,2,1,2,1}
-39\,\HBS_{2,2,2,1,1}
+29\,\HBS_{2,3,1,1,1}
-9\,\HBS_{3,1,1,1,2}
+8\,\HBS_{3,1,1,2,1}
+8\,\HBS_{3,1,2,1,1}\nonumber\\[1.12mm]&&\hspace*{-3mm}
+21\,\HBS_{3,2,1,1,1}
-16\,\HBS_{4,1,1,1,1}
+12\,\HBS_{1,1,1,2,2,1}
-18\,\HBS_{1,1,1,3,1,1}
+26\,\HBS_{1,1,2,2,1,1}\nonumber\\[1.12mm]&&\hspace*{-3mm}
-28\,\HBS_{1,1,3,1,1,1}
+24\,\HBS_{1,2,2,1,1,1}
-16\,\HBS_{1,3,1,1,1,1}
+16\,\HBS_{2,2,1,1,1,1}
-16\,\HBS_{3,1,1,1,1,1}
\big)
\,.\label{ABAz3}
\eeqa
We check with $\M=40$ that the obtained expression is indeed correct.
The final expression for the six-loop anomalous dimension of twist-two operators from ABA in the canonical basis of the usual harmonic sums (\ref{vhs}) can be found as the ancillary files of the arXiv version this paper or on the web-page \href{http://thd.pnpi.spb.ru/~velizh/6loop/}{\texttt{http://thd.pnpi.spb.ru/\textasciitilde velizh/6loop/}}.

\section{Calculations of the six-loop anomalous dimension for finite $M$} \label{sec:finiteS}
The quantum spectral curve \cite{Gromov:2013pga,Gromov:2014caa} formulates the spectral problem of planar $\cN=4$ SYM in terms of a finite set of Riemann-Hilbert equations. It was derived based on several assumptions, in particular that the theory is integrable. This approach automatically captures all the features of the spectral problem, i.e. no notion of ABA and wrapping contributions is needed. Recently, QSC was solved perturbatively for any operator in the $\mathfrak{sl}(2)$ sector by two of the authors \cite{Marboe:2014gma}, and an efficient \texttt{MATHEMATICA}-implementation of this iterative algorithm was provided, allowing up to 10-loop calculations of the simplest operators on a standard laptop. In our work, this procedure has been used to produce the six-loop anomalous dimensions of twist-two operators for finite $M$. Naturally, the computation time increases significantly with the value of $M$ (15 seconds for $M=2$ compared to five hours for $M=80$), since the algorithm is systematically operating with the Baxter polynomial which is a degree-$M$ polynomial. We refer to \cite{Marboe:2014gma} for a detailed description of the procedure as well as access to the \texttt{MATHEMATICA}-implementation.

To have a sufficient amount of finite $M$ results, we determined $\gamma_{12}(M)$ for the 40 lowest even integer spins, i.e. for $M=2,4,\hdots,80$.
We here provide the first five results obtained by the method. The remaining results can be reproduced from the general result below.
\begin{eqnarray}
\gamma_{12}(2) &=& 48(-160 - 5472 \,\zeta_3 - 432 \,\zeta_3^2 + 2340 \,\zeta_5 + 3240 \,\zeta_3 \,\zeta_5 + 1575 \,\zeta_7 - 10206 \,\zeta_9)
\,,\nonumber\\[1.53mm]
\gamma_{12}(4) &=& 25\left(\frac{493794415}{118098} -
\frac{76738930}{2187} \,\zeta_3 - \frac{545000}{81} \,\zeta_3^2
+ \frac{11977223}{729} \,\zeta_5+ \frac{50000}{3} \,\zeta_3
\,\zeta_5 \nonumber\right.\\&&\left. + \frac{1723925}{81} \,\zeta_7 - 37800 \,\zeta_9\right)\nonumber\,,\\[1.53mm]
\gamma_{12}(6)&=& \frac{49}{25} \left( \frac{1305379290116927483}{14762250000000} -
\frac{2757220598468}{3796875} \,\zeta_3 -
\frac{80404688}{375} \,\zeta_3^2\nonumber\right. \\ &&\left.+ \frac{1831031111}{5625}
\,\zeta_5 +345744 \,\zeta_3 \,\zeta_5 +
\frac{44325547}{75} \,\zeta_7 - 666792  \,\zeta_9\right)\nonumber\,,\\[1.53mm]
\gamma_{12}(8) &=&\frac{1}{175}\left( \frac{633898152590486916969605010931}{19062721117944000000000} -
\frac{125986766100891990916921}{400241898000000} \,\zeta_3\right. \nonumber\\&&-
\frac{467665307183261}{3601500} \,\zeta_3^2 +
\frac{2326437396820615637}{18151560000} \,\zeta_5 +
\frac{7932799458}{49} \,\zeta_3 \,\zeta_5 \nonumber\\&&\left.+
\frac{19149255965089}{58800} \,\zeta_7 - 281452806
\,\zeta_9\right)\nonumber\,,\\[1.53mm]
\gamma_{12}(10) &=& \frac{1331}{175} \bigg( \frac{474005292705148459163445336569621}{21612184690434724915200000000} \nonumber\\&&-
\frac{12417931006251790179790817}{47267767670004000000}
\,\zeta_3 - \frac{3452378452329719}{23629441500} \,\zeta_3^2 \nonumber\\&& +
\frac{6477832741701863069}{71455431096000} \,\zeta_5 + \frac{604223422}{3969} \,\zeta_3 \,\zeta_5 +
\frac{1757559752377}{5143824} \,\zeta_7 \nonumber\\&&
-245586\,\zeta_9\bigg)\,.
\end{eqnarray}
The first 35 values, i.e. $M\le 70$, were used to construct the general structure for any $M$, and it was then checked that the last five results, $72\le M\le 80$, are consistent with this structure.

\section{Reconstruction of the general form of six-loop anomalous dimension} \label{sec:L6}

In this section we describe in detail the method of the reconstruction of the anomalous dimension of twist-two operators from the above results and special numerical algorithms. It appears to be suitable to compute the wrapping correction part separately, without the ABA part, and then add it to the ABA part found in section~\ref{sec:ABA}. Similarly to the ABA case, we will look for the reciprocity function $\reciP^{\Wrap}$ which is represented in a smaller basis of binomial harmonic sums \eqref{BinomialSums} and then reconstruct the conformal dimension using \eqref{PABAExp}.

Let us start by recalling  how a simpler five-loop computation is done. At five loops, $\gamma_{10}^{\Wrap}$ and $\reciP_{10}^{\Wrap}$ are related as
\beq \label{gamma10P}
\gamma_{10}^{\Wrap}=
\frac{1}{2} \Big(\gamma_{2} \gamma^{\Wrap}_{8}\Big)'+\cP_{10}^{\Wrap}\,,
\eeq
where $\gamma_{2}$ and $\gamma^{\Wrap}_{8}$ are given by \cite{Bajnok:2008qj}
\beqa
&&\gamma^{\Wrap}_{8}=\reciP_{8}^{\Wrap}=\reciP_{2}^{2}\,\cT_{8}\,,\\
&&\cT_{8}=\Big(-5\,\z5 +2\,\HBS_{2}\,\z3+\left(\HBS_{2,1,2}-\HBS_{3,1,1}\right)\Big)\,,\\
&&\gamma_{2}=\cP_{2}=4\,\HBS_1\,.
\eeqa

To find $\cP_{10}^{\Wrap}$ one could naively attempt to consider the full basis of binomial harmonic sums of given \transcendentalitylevel. However, it is known, from a structure analysis of the L\"{u}scher formulae, that a simpler ansatz should work~\cite{Lukowski:2009ce}:
\beqa
\cP_{10}^{\Wrap}&=&
\cP_{2}^2\, \cT_{10}
+\left(c_1\,\cP_{2} \cP_{4} +c_2\,\cP_{2}^4\right)\, \cT_{8}\,,
\label{P10Ansatz}\\
\cP_{4}&=&8\,\big(\HBS_1\HBS_2-\HBS_{2,1}-\HBS_3\big)\,.\label{P4}
\eeqa

During the computations of the wrapping corrections at five loops this ansatz was assumed in advance and this allowed finding the final result much faster. Indeed, one found in ref.~\cite{Lukowski:2009ce}
\beqa \label{gamma10split}
\cP_{10}^{\Wrap}&=&2\,\cP_2^2 \cT_{10}
+2\,\cP_2 \Big(2\,\cP_4 +\frac{1}{16} \cP^3_2\Big)
\Big(-5\,\z5 +2\,\HBS_{2}\,\z3+\left(\HBS_{2,1,2}-\HBS_{3,1,1}\right)\Big)\,,\\[2mm] \label{Ttilde}
\cT_{10}&=&
105\,\z7
-6\,\HBS_1\,\zp{3}{2}
-40\,\HBS_2\,\z5
+4\big( 3\,\HBS_1 \HBS_{2,1} -2\,\HBS_{2,2}+2\,\HBS_{3,1}-\HBS_{2,1,1}-\HBS_4\big)\,\z3\nonumber\\
&+&2 \, \Big(
\HBS_1 \left(\HBS_{2,3,1}-\HBS_{3,1,2}\right)
-\HBS_{2,1,4}
+2\, \HBS_{2,2,3}
-5\, \HBS_{3,1,3}
+2\, \HBS_{3,2,2}\nonumber\\
&&\quad\ +2\, \HBS_{3,3,1}
-\HBS_{4,1,2}
+\HBS_{5,1,1}
-2\, \HBS_{2,1,2,2}
+2\, \HBS_{2,1,3,1}
-2\, \HBS_{2,2,1,2}
-2\, \HBS_{2,2,2,1}\nonumber\\
&&\quad\ +2\, \HBS_{2,3,1,1}
-2\, \HBS_{3,1,1,2}
+2\, \HBS_{3,1,2,1}
+2\, \HBS_{3,2,1,1}
-\HBS_{2,1,1,1,2}
+\HBS_{3,1,1,1,1}\Big)\,.
\eeqa

The advantage of the reconstruction of the wrapping correction part instead of the full anomalous dimension is clear from eq.~(\ref{P10Ansatz}): $\cT_{10}$ multiplied by $\HBS_1^2$ has the tran\-scen\-den\-tality~7 with $2^{9-1-2}=64$ binomial harmonic sums in the basis instead of the transcendentality~9 with $2^{9-1}=256$ binomial harmonic sums in the basis for $\cP_{10}^{\Wrap}$, so we will have considerably less binomial harmonic sums in the ansatz.

At five loops, one could imagine to not rely on \eqref{P10Ansatz} and do a brute-force approach for fixing about 300 coefficients in the most general ansatz respecting transcendentality. However, brute-forcing is likely to be impossible at six loops. Indeed, one has to fix more than 1000 coefficients if no further restrictions apply which is currently beyond the practical computational ability of the QSC approach. Hence one should attempt to generalise \eqref{P10Ansatz} to the six-loop case.

At six loops $\gamma_{12}^{\Wrap}$ and $\reciP_{12}^{\Wrap}$ are related as
\beq \label{gamma12P}
\gamma_{12}^{\Wrap}=
\frac{1}{2} \Big(\gamma_4 \gamma^{\Wrap}_{8}+\gamma_2 \gamma^{\Wrap}_{10}\Big)'
-\frac{1}{8} \Big(\gamma_2^2 \gamma^{\Wrap}_{8}\Big)''
+\cP_{12}^{\Wrap}\,.
\eeq

A straightforward guess generalising eq.~(\ref{P10Ansatz}) would be the following one
\beqa
\cP_{12}^{\Wrap}&=&
\cP_2^2\, \cT_{12}
+\Big(c_1\,\cP_2 \cP_4
+c_2\,\cP_2^4\Big) \cT_{10}
+\Big(c_3\,\cP_2 \cP_6
+c_4\,\cP_2^3 \cP_4
+c_5\,\cP_2^6
+c_6\,\cP_4^2\Big) \cT_{8}\,,
\ \qquad\label{P12Ansatz}\\
&=&
\cP_{12,\,{\mathrm{rational}}}^{\Wrap}
+\z3\cP_{12,\,\z3}^{\Wrap}
+\z5\cP_{12,\,\z5}^{\Wrap}
+\zp{3}{2}\cP_{12,\,\zp{3}{2}}^{\Wrap}
+\z7\cP_{12,\,\z7}^{\Wrap}
+\z3\z5\cP_{12,\,\z3\z5}^{\Wrap}\nonumber\\&&\hspace*{20mm}
+\zp{3}{3}\cP_{12,\,\zp{3}{3}}^{\Wrap}
+\z9\cP_{12,\,\z9}^{\Wrap}\label{P12AnsatzZs}
\eeqa
with $\cT_{12}$ and $\reciP_{6}$ given by
\beqa
\cT_{12}&=&
\z9\, \reciT_{\z9}
+\zp{3}{3}\, \reciT_{\zp{3}{3}}
+\z5\,\z3\, \reciT_{\z5\z3}
+\z7\, \reciT_{\z7}
+\zp{3}{2}\, \reciT_{\zp{3}{2}}
+\z5\, \reciT_{\z5}
+\z3\, \reciT_{\z3}
+\reciT_{\mathrm{rational}}\,,\ \qquad\label{reciT}\\
\cP_{6}&=&16
\Big(\HBS_1^2\HBS_3
+\HBS_1
\big(
\HBS_{3,1}
-\HBS_{2,2}
-4\, \HBS_4
\big)
+2 \big(
\HBS_{2,2,1}
-\HBS_{2,1,2}
-\HBS_{3,2}
+3\, \HBS_5
\big)
\Big)
\,.\label{P6}
\eeqa
The idea is that the genuine 6-loop wrapping term $\cP_2^2\, \cT_{12}$ should be a multiple of $\cP_2^2$, as follows from a basic analysis of L\"{u}scher formulae, whereas all other terms in $\cP_{12}^{\Wrap}$ are constructed from the lower-loop expressions. Keeping in mind that $\cT_{2n}=0$ for $n<4$, we wrote down the most general possible combinations assuming that the expressions in brackets are expressible in terms of $\cP_n$, similarly to what happened at five loops.

Applying the principle of maximal transcendentality we conclude that the transcendentality of the components
$\cT_{\z9}$, $\cT_{\zp{3}{3}}$, $\cT_{\z3\z5}$, $\cT_{\z7}$, $\cT_{\zp{3}{2}}$, $\cT_{\z5}$, $\cT_{\z3}$, $\cT_{\rm rational}$ should be equal to 0, 0, 1, 2, 3, 4, 6 and 9 respectively (see Table~\ref{table:nums} for the numbers of the binomial harmonic sums in the corresponding ansatz). The lowest-transcendentality functions,
$\cT_{\z9}$, $\cT_{\zp{3}{3}}$, $\cT_{\z3\z5}$, $\cT_{\z7}$, $\cT_{\zp{3}{2}}$, $\cT_{\z5}$ may be obtained exactly by solving the equations for the corresponding contribution as we know the full result for even values of $\M$ up to $\M=70$ (but we should take into account that the $\z7$ contribution is contained in $\cT_{10}$, while $\z5$ and $\z3$ contributions are contained in both $\cT_{8}$ and $\cT_{10}$). Indeed, from the knowledge of just the Konishi operator ($M=2$) one can find
\beqa
&&\cT_{\z9}=-3402\,,\label{T12z9}\\[2mm]
&&\cT_{\zp{3}{3}}=0\,,\label{T12z33}\\[2mm]
&&\cT_{\z3\z5}=360\,\label{T12z3z5}\HBS_1\,,
\eeqa
and using the first few values one can easily obtain
\beqa
&&\cP_{12,\,\zp{3}{2}}^{\Wrap}=-12\, \cP_2^2 \Big(\HBS_1^3+14\, \HBS_1 \HBS_2-6 \big(\HBS_{2,1}+\HBS_3\big)\Big)\,.\label{Resultz32}
\eeqa
However, already for $\cT_{\z7}$ we cannot find any solution, despite that the possible ansatz contains only two binomial harmonic sums: $\HBS_2$ and $\HBS_1^2$. To find a correct expression for $\zeta_7$-term, we remove the restriction of eq.~(\ref{P12Ansatz}), that is we take the ansatz with all possible binomial harmonic sums with transcendentality 4:
\beqa
{\mathrm{{Basis}}}\big[\cP^{\Wrap}_{12,\,\z7}\big]&=&\Big\{\HBS_4,\,\HBS_{3,1},\,\HBS_{2,2},\,\HBS_{2,1,1},\,\HBS_1\HBS_3,\,\HBS_1\HBS_{2,1},\,\HBS_1^2\HBS_2,\,\HBS_1^4 \Big\}.
\eeqa
For this most general ansatz a solution exists (note that we have 35 equations for 8 variables):
\beqa
\cP^{\Wrap}_{12,\,\z7}&=&
224\,\HBS_1 \Big(12\, \HBS_1^3+151\, \HBS_1 \HBS_2-72\, \HBS_3-52\, \HBS_{2,1}\Big)\,.\label{Resultz7}
\eeqa
Comparing this result to $\cP_{4}$ from eq.~(\ref{P4}) one can see that the coefficients of $\HBS_1\HBS_3$ and $\HBS_1\HBS_{2,1}$ are different, so our simple assumption about a general form of $\cP_{12}^{\Wrap}$ is not correct and we should extend all bases for $\z5$, $\z3$ and rational contributions to $\cP_{12}^{\Wrap}$. We have found that a minimal extension can be performed by expanding all terms with the brackets in eq.~(\ref{P12Ansatz}), that is we should expand
$\cP_2 \cP_6 \cT_{8}$,
$\cP_2^3 \cP_4 \cT_{8}$,
$\cP_2^6 \cT_{8}$,
$\cP_4^2 \cT_{8}$,
$\cP_2 \cP_4 \cT_{10}$,
$\cP_2^4 \cT_{10}$ and add all possible terms from these expansions to the corresponding ansatz for $\cP_{12,\z5}^{\Wrap}$, $\cP_{12,\z3}^{\Wrap}$ or $\cP_{12,\mathrm{rational}}^{\Wrap}$ contributions.
The ansatz for the $\z5$ contribution to $\cP_{12}^{\Wrap}$ will contain the following binomial harmonic sums:
\beqa
{\mathrm{{Basis}}}\big[\cP^{\Wrap}_{12,\,\z5}\big]&=&\Big\{\HBS_1^2\HBS_4,\,\HBS_1^2\HBS_{3,1},\,\HBS_1^2\HBS_{2,2},\,\HBS_1^2\HBS_{2,1,1},\,\HBS_1^3\HBS_3,\,\HBS_1^3\HBS_{2,1},\,
\HBS_1^4\HBS_2,\,\HBS_1^6,\nonumber\\
&&\hspace*{-10mm}\HBS_1\HBS_2\HBS_3,\,\HBS_1\HBS_2\HBS_{2,1},\,\HBS_1\HBS_{2,2,1},\,\HBS_1\HBS_{2,1,2},\,\HBS_1\HBS_{3,2},\,\HBS_1\HBS_{5},
\big(\HBS_1\HBS_{2}-\HBS_3-\HBS_{2,1}\big)^2\Big\},\qquad\label{Ansatzz5}
\eeqa
where the first line corresponds to the basis for $\cT_{\z5}$ while the second line comes from the expansion of eq.~(\ref{P12Ansatz}). Note that for the $\z5$ contribution we have found that the last term in eq.~(\ref{Ansatzz5}), which corresponds to $\cP_{4}^2$, does not need to be expanded and this property is also correct for the other contributions. So we have 16 binomial harmonic sums in the ansatz for the $\z5$ contribution and the coefficients of these sums can be found from the known values of the anomalous dimension at fixed $M$. The final result for the $\z5$ contribution is given by:
\beqa
\cP^{\Wrap}_{12,\,\z5}&=&
8 \Big(
\frac{2944}{3} \HBS_1 \HBS_2 \HBS_{2,1}
-\frac{1760}{3} \HBS_1^3 \HBS_{2,1}
-\frac{436}{3} \HBS_1^4 \HBS_2
-\frac{25}{3}\HBS_1^6
-80 \big(-\HBS_{2,1}+\HBS_1 \HBS_2-\HBS_3\big)^2\nonumber\\[2mm]&&\quad\
-\,160\, \HBS_1^3 \HBS_3
+1024\, \HBS_1^2 \HBS_4
+656\, \HBS_1^2 \HBS_{2,2}
-2024\, \HBS_1^2 \HBS_{3,1}
+320\, \HBS_1^2 \HBS_{2,1,1}\nonumber\\[2mm]&&\quad\
-\,320\, \HBS_1 \HBS_{2,2,1}
+608\, \HBS_1\HBS_2 \HBS_3
-960\, \HBS_1 \HBS_5
\Big).\label{Resultz5}
\eeqa

For the $\z3$ contribution our minimal ansatz consists of $2^{8-1-2}+19=51$ binomial harmonic sums, while we have only 35 values. To find the coefficients in the ansatz we use the same method as we used for the reconstruction of the $\z3$ contribution to the ABA part of the anomalous dimension in eq.~(\ref{ABAz3}).
With the help of the \texttt{LatticeReduce} function from \texttt{MATHEMATICA}, which realizes the \texttt{LLL}-algorithm, we have found the following general expression for the $\z3$ contribution to $\cP_{12}^{\Wrap}$:
\beqa
\cP^{\Wrap}_{12,\,\z3}&=&
32\, \bigg(
\frac{5}{6} \HBS_1^6 \HBS_2
+12\, \HBS_1^5 \HBS_{2,1}
-\frac{10}{3} \HBS_1^4 \HBS_{2,1,1}
+\frac{46}{3} \HBS_1^4 \HBS_{2,2}
+\frac{49}{3} \HBS_1^4 \HBS_2^2
-\frac{10}{3} \HBS_1^4 \HBS_4\nonumber\\[1.12mm]&&\hspace*{-9mm}
+\,\frac{248}{3} \HBS_1^3\HBS_2  \HBS_{2,1}
+8\,  \HBS_1^3 \HBS_2 \HBS_3
-8\, \HBS_1^3 \HBS_{2,3}
+16\, \HBS_1^3 \HBS_{3,2}
-16\, \HBS_1^3 \HBS_{4,1}
-80\, \HBS_1^3 \HBS_{2,2,1}
-24\, \HBS_1^3 \HBS_{2,1,1,1}\nonumber\\[1.12mm]&&\hspace*{-9mm}
-\,66\, \HBS_1^2 \HBS_{2,1}^2
-\frac{332}{3} \HBS_1^2\HBS_3  \HBS_{2,1}
-66\, \HBS_1^2 \HBS_2 \HBS_{2,2}
+\frac{82}{3} \HBS_1^2 \HBS_2^3
-118\, \HBS_1^2 \HBS_2 \HBS_4
+20\, \HBS_1^2 \HBS_6
+12\, \HBS_1^2 \HBS_{2,4}\nonumber\\[1.12mm]&&\hspace*{-9mm}
+\,\frac{116}{3} \HBS_1^2 \HBS_{3,3}
+\frac{152}{3} \HBS_1^2 \HBS_{4,2}
+22\, \HBS_1^2 \HBS_{5,1}
-32\, \HBS_1^2 \HBS_2 \HBS_{2,1,1}
+\frac{92}{3} \HBS_1^2 \HBS_{2,1,3}
-22\, \HBS_1^2 \HBS_{2,3,1}\nonumber\\[1.12mm]&&\hspace*{-9mm}
+\,14\, \HBS_1^2 \HBS_{4,1,1}
+34\, \HBS_1^2 \HBS_{2,2,1,1}
+20\, \HBS_1^2 \HBS_{2,1,1,1,1}
+24\, \HBS_1 \HBS_2^2 \HBS_3
+96\, \HBS_1 \HBS_2 \HBS_5
-48\, \HBS_1 \HBS_2^2  \HBS_{2,1}\nonumber\\[1.12mm]&&\hspace*{-9mm}
-\,48\, \HBS_1 \HBS_3 \HBS_{2,2}
+32\, \HBS_1 \HBS_{2,1} \HBS_{2,2}
+32\, \HBS_1 \HBS_2 \HBS_{2,2,1}
+32\, \HBS_1 \HBS_3 \big(2\, \HBS_{2,2} -2\, \HBS_{3,1} + \HBS_{2,1,1} + \HBS_4\big)\nonumber\\[1.12mm]&&\hspace*{-9mm}
+\,\frac{16}{3} \HBS_1 \HBS_{2,1} \big(2\, \HBS_{2,2} - 2\, \HBS_{3,1} + \HBS_{2,1,1} + \HBS_4\big)
+8\, \HBS_2 \big(\HBS_1 \HBS_2-\HBS_{2,1} - \HBS_3\big)^2
\bigg).\label{Resultz3}
\eeqa

The last part is the rational contribution.
In this case our minimal ansatz consists of $2^{11-1-2}+67=323$ binomial harmonic sums. Nevertheless the available 35 values are enough to find all coefficients in this ansatz with the help of the \texttt{LLL}-algorithm, but for this purpose we used a \texttt{C++} implementation of the \texttt{LLL}-algorithm in the form of the \texttt{fplll}-program~\cite{fplll}. After about one hour of computations we obtained the following general expression for the rational contribution to $\cP_{12}^{\Wrap}$:
\beqa
\cP^{\Wrap}_{12,\,{\mathrm{rational}}}&=&
-\frac{32}{3}\Bigg(
\frac{5}{4}\,\HBS_1^6 \HBS_{3,1,1}
-\frac{5}{4}\,\HBS_1^6 \HBS_{2,1,2}
-6\,\HBS_1^5 \HBS_{2,3,1}
+6\,\HBS_1^5 \HBS_{3,1,2}
-25\,\HBS_1^4 \HBS_2\HBS_{2,1,2}
+5\,\HBS_1^4 \HBS_{2,1,4}\nonumber\\[1.51mm]&&\hspace*{-17.9mm}
-4\,\HBS_1^4 \HBS_{2,2,3}
+33\,\HBS_1^4 \HBS_2\HBS_{3,1,1}
+7\,\HBS_1^4 \HBS_{3,1,3}
-4\,\HBS_1^4 \HBS_{3,2,2}
-12\,\HBS_1^4 \HBS_{3,3,1}
+5\,\HBS_1^4 \HBS_{4,1,2}
-5\,\HBS_1^4 \HBS_{5,1,1}\nonumber\\[1.51mm]&&\hspace*{-17.9mm}
+13\,\HBS_1^4 \HBS_{2,1,2,2}
+\,\HBS_1^4 \HBS_{2,1,3,1}
+\,\HBS_1^4 \HBS_{2,2,1,2}
+12\,\HBS_1^4 \HBS_{2,2,2,1}
-13\,\HBS_1^4 \HBS_{2,3,1,1}
+4\,\HBS_1^4 \HBS_{3,1,1,2}\nonumber\\[1.51mm]&&\hspace*{-17.9mm}
-13\,\HBS_1^4 \HBS_{3,1,2,1}
-13\,\HBS_1^4 \HBS_{3,2,1,1}
+5\,\HBS_1^4 \HBS_{2,1,1,1,2}
-5\,\HBS_1^4 \HBS_{3,1,1,1,1}
-27\,\HBS_1^3 \HBS_3\HBS_{2,1,2}\nonumber\\[1.51mm]&&\hspace*{-17.9mm}
+10\,\HBS_1^3 \HBS_{2,1}\HBS_{2,1,2}
+2\,\HBS_1^3 \HBS_{2,2,4}
-55\,\HBS_1^3 \HBS_2\HBS_{2,3,1}
+12\,\HBS_1^3 \HBS_{2,5,1}
+37\,\HBS_1^3 \HBS_3\HBS_{3,1,1}\nonumber\\[1.51mm]&&\hspace*{-17.9mm}
-18\,\HBS_1^3 \HBS_{2,1}\HBS_{3,1,1}
+54\,\HBS_1^3 \HBS_2\HBS_{3,1,2}
-18\,\HBS_1^3 \HBS_{3,1,4}
-21\,\HBS_1^3 \HBS_{3,2,3}
+15\,\HBS_1^3 \HBS_{3,4,1}
-13\,\HBS_1^3 \HBS_{4,1,3}\nonumber\\[1.51mm]&&\hspace*{-17.9mm}
+15\,\HBS_1^3 \HBS_{4,3,1}
-12\,\HBS_1^3 \HBS_{5,1,2}
+22\,\HBS_1^3 \HBS_{2,1,3,2}
+15\,\HBS_1^3 \HBS_{2,1,4,1}
-23\,\HBS_1^3 \HBS_{2,2,1,3}
+6\,\HBS_1^3 \HBS_{2,2,2,2}\nonumber\\[1.51mm]&&\hspace*{-17.9mm}
+51\,\HBS_1^3 \HBS_{2,2,3,1}
-10\,\HBS_1^3 \HBS_{2,3,1,2}
+24\,\HBS_1^3 \HBS_{2,3,2,1}
-27\,\HBS_1^3 \HBS_{3,1,1,3}
-45\,\HBS_1^3 \HBS_{3,1,2,2}
+24\,\HBS_1^3 \HBS_{3,1,3,1}\nonumber\\[1.51mm]&&\hspace*{-17.9mm}
-29\,\HBS_1^3 \HBS_{3,2,1,2}
-7\,\HBS_1^3 \HBS_{3,2,2,1}
+\,\HBS_1^3 \HBS_{3,3,1,1}
+12\,\HBS_1^3 \HBS_{2,1,1,3,1}
-4\,\HBS_1^3 \HBS_{2,1,2,1,2}
+33\,\HBS_1^3 \HBS_{2,1,2,2,1}\nonumber\\[1.51mm]&&\hspace*{-17.9mm}
+4\,\HBS_1^3 \HBS_{2,1,3,1,1}
-\,\HBS_1^3 \HBS_{2,2,2,1,1}
+12\,\HBS_1^3 \HBS_{2,3,1,1,1}
-12\,\HBS_1^3 \HBS_{3,1,1,1,2}
-14\,\HBS_1^3 \HBS_{3,1,1,2,1}\nonumber\\[1.51mm]&&\hspace*{-17.9mm}
-14\,\HBS_1^3 \HBS_{3,1,2,1,1}
+124\,\HBS_1^2 \HBS_4\HBS_{2,1,2}
+37\,\HBS_1^2 \HBS_{2,2}\HBS_{2,1,2}
-48\,\HBS_1^2 \HBS_{3,1}\HBS_{2,1,2}
+13\,\HBS_1^2 \HBS_{2,1,1}\HBS_{2,1,2}\nonumber\\[1.51mm]&&\hspace*{-17.9mm}
+45\,\HBS_1^2 \HBS_2\HBS_{2,1,4}
-30\,\HBS_1^2 \HBS_{2,1,6}
-25\,\HBS_1^2 \HBS_2\HBS_{2,2,3}
+16\,\HBS_1^2 \HBS_{2,2,5}
-17\,\HBS_1^2 \HBS_3\HBS_{2,3,1}
+53\,\HBS_1^2 \HBS_{2,1}\HBS_{2,3,1}\nonumber\\[1.51mm]&&\hspace*{-17.9mm}
-150\,\HBS_1^2 \HBS_4\HBS_{3,1,1}
-66\,\HBS_1^2 \HBS_{2,2}\HBS_{3,1,1}
+67\,\HBS_1^2 \HBS_{3,1}\HBS_{3,1,1}
+5\,\HBS_1^2 \HBS_{2,1,1}\HBS_{3,1,1}
+25\,\HBS_1^2 \HBS_3\HBS_{3,1,2}\nonumber\\[1.51mm]&&\hspace*{-17.9mm}
-50\,\HBS_1^2 \HBS_{2,1}\HBS_{3,1,2}
+29\,\HBS_1^2 \HBS_2\HBS_{3,1,3}
-18\,\HBS_1^2 \HBS_{3,1,5}
+10\,\HBS_1^2 \HBS_2\HBS_{3,2,2}
+52\,\HBS_1^2 \HBS_{3,2,4}
-57\,\HBS_1^2 \HBS_2\HBS_{3,3,1}\nonumber\\[1.51mm]&&\hspace*{-17.9mm}
+24\,\HBS_1^2 \HBS_{3,5,1}
+15\,\HBS_1^2 \HBS_2\HBS_{4,1,2}
+30\,\HBS_1^2 \HBS_{4,1,4}
-10\,\HBS_1^2 \HBS_{4,2,3}
-4\,\HBS_1^2 \HBS_{4,4,1}
-22\,\HBS_1^2 \HBS_2\HBS_{5,1,1}\nonumber\\[1.51mm]&&\hspace*{-17.9mm}
+20\,\HBS_1^2 \HBS_{5,1,3}
+20\,\HBS_1^2 \HBS_{5,2,2}
-4\,\HBS_1^2 \HBS_{5,3,1}
-30\,\HBS_1^2 \HBS_{6,1,2}
+30\,\HBS_1^2 \HBS_{7,1,1}
+52\,\HBS_1^2 \HBS_2\HBS_{2,1,2,2}\nonumber\\[1.51mm]&&\hspace*{-17.9mm}
-18\,\HBS_1^2 \HBS_{2,1,2,4}
-23\,\HBS_1^2 \HBS_2\HBS_{2,1,3,1}
-31\,\HBS_1^2 \HBS_{2,1,3,3}
-18\,\HBS_1^2 \HBS_{2,1,4,2}
+3\,\HBS_1^2 \HBS_{2,1,5,1}
+86\,\HBS_1^2 \HBS_2\HBS_{2,2,1,2}\nonumber\\[1.51mm]&&\hspace*{-17.9mm}
+18\,\HBS_1^2 \HBS_{2,2,1,4}
-4\,\HBS_1^2 \HBS_2\HBS_{2,2,2,1}
-\,\HBS_1^2 \HBS_{2,2,2,3}
+10\,\HBS_1^2 \HBS_{2,2,3,2}
-44\,\HBS_1^2 \HBS_{2,2,4,1}
-47\,\HBS_1^2 \HBS_2\HBS_{2,3,1,1}\nonumber\\[1.51mm]&&\hspace*{-17.9mm}
-4\,\HBS_1^2 \HBS_{2,3,1,3}
+14\,\HBS_1^2 \HBS_{2,3,2,2}
+6\,\HBS_1^2 \HBS_{2,3,3,1}
-18\,\HBS_1^2 \HBS_{2,4,1,2}
+18\,\HBS_1^2 \HBS_{2,5,1,1}
+25\,\HBS_1^2 \HBS_2\HBS_{3,1,1,2}\nonumber\\[1.51mm]&&\hspace*{-17.9mm}
+44\,\HBS_1^2 \HBS_{3,1,1,4}
-62\,\HBS_1^2 \HBS_2\HBS_{3,1,2,1}
+2\,\HBS_1^2 \HBS_{3,1,3,2}
-14\,\HBS_1^2 \HBS_{3,1,4,1}
-52\,\HBS_1^2 \HBS_2\HBS_{3,2,1,1}
-14\,\HBS_1^2 \HBS_{3,2,1,3}\nonumber\\[1.51mm]&&\hspace*{-17.9mm}
+30\,\HBS_1^2 \HBS_{3,2,2,2}
-13\,\HBS_1^2 \HBS_{3,2,3,1}
-5\,\HBS_1^2 \HBS_{3,3,1,2}
+23\,\HBS_1^2 \HBS_{3,3,2,1}
+9\,\HBS_1^2 \HBS_{3,4,1,1}
+10\,\HBS_1^2 \HBS_{4,1,1,3}\nonumber\\[1.51mm]&&\hspace*{-17.9mm}
-48\,\HBS_1^2 \HBS_{4,1,2,2}
+2\,\HBS_1^2 \HBS_{4,1,3,1}
+35\,\HBS_1^2 \HBS_{4,2,1,2}
-23\,\HBS_1^2 \HBS_{4,2,2,1}
+14\,\HBS_1^2 \HBS_{4,3,1,1}
-20\,\HBS_1^2 \HBS_{5,1,1,2}\nonumber\\[1.51mm]&&\hspace*{-17.9mm}
+18\,\HBS_1^2 \HBS_{5,1,2,1}
+18\,\HBS_1^2 \HBS_{5,2,1,1}
+38\,\HBS_1^2 \HBS_2\HBS_{2,1,1,1,2}
-6\,\HBS_1^2 \HBS_{2,1,1,1,4}
+4\,\HBS_1^2 \HBS_{2,1,1,4,1}
-2\,\HBS_1^2 \HBS_{2,1,2,1,3}\nonumber\\[1.51mm]&&\hspace*{-17.9mm}
-27\,\HBS_1^2 \HBS_{2,1,2,2,2}
+22\,\HBS_1^2 \HBS_{2,1,2,3,1}
-25\,\HBS_1^2 \HBS_{2,1,3,1,2}
+7\,\HBS_1^2 \HBS_{2,1,3,2,1}
-12\,\HBS_1^2 \HBS_{2,1,4,1,1}\nonumber 
\eeqa
\beqa&&\hspace*{3.2mm}
+38\,\HBS_1^2 \HBS_{2,2,1,1,3}
-60\,\HBS_1^2 \HBS_{2,2,1,2,2}
+48\,\HBS_1^2 \HBS_{2,2,1,3,1}
-34\,\HBS_1^2 \HBS_{2,2,2,1,2}
-2\,\HBS_1^2 \HBS_{2,2,2,2,1}\nonumber\\[2mm]&&\hspace*{3.2mm}
-6\,\HBS_1^2 \HBS_{2,2,3,1,1}
-14\,\HBS_1^2 \HBS_{2,3,1,1,2}
+25\,\HBS_1^2 \HBS_{2,3,1,2,1}
+42\,\HBS_1^2 \HBS_{2,3,2,1,1}
-56\,\HBS_1^2 \HBS_2\HBS_{3,1,1,1,1}\nonumber\\[2mm]&&\hspace*{3.2mm}
-8\,\HBS_1^2 \HBS_{3,1,1,1,3}
-20\,\HBS_1^2 \HBS_{3,1,1,2,2}
+13\,\HBS_1^2 \HBS_{3,1,1,3,1}
-26\,\HBS_1^2 \HBS_{3,1,2,1,2}
+9\,\HBS_1^2 \HBS_{3,1,2,2,1}\nonumber\\[2mm]&&\hspace*{3.2mm}
+18\,\HBS_1^2 \HBS_{3,1,3,1,1}
-9\,\HBS_1^2 \HBS_{3,2,1,1,2}
+42\,\HBS_1^2 \HBS_{3,2,1,2,1}
+23\,\HBS_1^2 \HBS_{3,2,2,1,1}
+7\,\HBS_1^2 \HBS_{3,3,1,1,1}\nonumber\\[2mm]&&\hspace*{3.2mm}
-6\,\HBS_1^2 \HBS_{4,1,1,1,2}
+6\,\HBS_1^2 \HBS_{5,1,1,1,1}
-18\,\HBS_1^2 \HBS_{2,1,1,1,2,2}
-10\,\HBS_1^2 \HBS_{2,1,1,1,3,1}
-7\,\HBS_1^2 \HBS_{2,1,1,2,1,2}\nonumber\\[2mm]&&\hspace*{3.2mm}
-4\,\HBS_1^2 \HBS_{2,1,1,2,2,1}
-11\,\HBS_1^2 \HBS_{2,1,1,3,1,1}
-20\,\HBS_1^2 \HBS_{2,1,2,1,1,2}
-36\,\HBS_1^2 \HBS_{2,1,2,2,1,1}
+6\,\HBS_1^2 \HBS_{2,1,3,1,1,1}\nonumber\\[2mm]&&\hspace*{3.2mm}
-18\,\HBS_1^2 \HBS_{2,2,1,1,1,2}
-7\,\HBS_1^2 \HBS_{2,2,2,1,1,1}
+18\,\HBS_1^2 \HBS_{2,3,1,1,1,1}
+14\,\HBS_1^2 \HBS_{3,1,1,1,1,2}
+18\,\HBS_1^2 \HBS_{3,1,1,1,2,1}\nonumber\\[2mm]&&\hspace*{3.2mm}
+13\,\HBS_1^2 \HBS_{3,1,1,2,1,1}
+8\,\HBS_1^2 \HBS_{3,1,2,1,1,1}
+18\,\HBS_1^2 \HBS_{3,2,1,1,1,1}
-30\,\HBS_1^2 \HBS_{2,1,1,1,1,1,2}
+30\,\HBS_1^2 \HBS_{3,1,1,1,1,1,1}\nonumber\\[2mm]&&\hspace*{3.2mm}
+68\,\HBS_1 \HBS_{2,1,2}^2
-144\,\HBS_1 \HBS_5\HBS_{2,1,2}
+68\,\HBS_1 \HBS_{3,2}\HBS_{2,1,2}
-48\,\HBS_1 \HBS_3\HBS_{2,1,4}
-8\,\HBS_1 \HBS_{2,1}\HBS_{2,1,4}\nonumber\\[2mm]&&\hspace*{3.2mm}
-48\,\HBS_1 \HBS_{2,1,2}\HBS_{2,2,1}
+24\,\HBS_1 \HBS_3\HBS_{2,2,3}
+18\,\HBS_1 \HBS_{2,1}\HBS_{2,2,3}
+144\,\HBS_1 \HBS_5\HBS_{3,1,1}
-68\,\HBS_1 \HBS_{3,2}\HBS_{3,1,1}\nonumber\\[2mm]&&\hspace*{3.2mm}
-68\,\HBS_1 \HBS_{2,1,2}\HBS_{3,1,1}
+48\,\HBS_1 \HBS_{2,2,1}\HBS_{3,1,1}
-24\,\HBS_1 \HBS_3\HBS_{3,1,3}
-46\,\HBS_1 \HBS_{2,1}\HBS_{3,1,3}
+24\,\HBS_1 \HBS_3\HBS_{3,2,2}\nonumber\\[2mm]&&\hspace*{3.2mm}
-8\,\HBS_1 \HBS_{2,1}\HBS_{3,2,2}
-2\,\HBS_1 \HBS_3\HBS_{3,3,1}
+44\,\HBS_1 \HBS_{2,1}\HBS_{3,3,1}
-48\,\HBS_1 \HBS_3\HBS_{4,1,2}
+18\,\HBS_1 \HBS_{2,1}\HBS_{4,1,2}\nonumber\\[2mm]&&\hspace*{3.2mm}
+48\,\HBS_1 \HBS_3\HBS_{5,1,1}
+8\,\HBS_1 \HBS_{2,1}\HBS_{5,1,1}
-24\,\HBS_1 \HBS_3\HBS_{2,1,2,2}
-64\,\HBS_1 \HBS_{2,1}\HBS_{2,1,2,2}
-2\,\HBS_1 \HBS_3\HBS_{2,1,3,1}\nonumber\\[2mm]&&\hspace*{3.2mm}
+24\,\HBS_1 \HBS_{2,1}\HBS_{2,1,3,1}
-24\,\HBS_1 \HBS_3\HBS_{2,2,1,2}
-90\,\HBS_1 \HBS_{2,1}\HBS_{2,2,1,2}
+2\,\HBS_1 \HBS_3\HBS_{2,2,2,1}
-44\,\HBS_1 \HBS_{2,1}\HBS_{2,2,2,1}\nonumber\\[2mm]&&\hspace*{3.2mm}
+24\,\HBS_1 \HBS_3\HBS_{2,3,1,1}
+64\,\HBS_1 \HBS_{2,1}\HBS_{2,3,1,1}
-24\,\HBS_1 \HBS_3\HBS_{3,1,1,2}
+8\,\HBS_1 \HBS_{2,1}\HBS_{3,1,1,2}
-2\,\HBS_1 \HBS_3\HBS_{3,1,2,1}\nonumber\\[2mm]&&\hspace*{3.2mm}
+64\,\HBS_1 \HBS_{2,1}\HBS_{3,1,2,1}
+24\,\HBS_1 \HBS_3\HBS_{3,2,1,1}
+64\,\HBS_1 \HBS_{2,1}\HBS_{3,2,1,1}
-48\,\HBS_1 \HBS_3\HBS_{2,1,1,1,2}\nonumber\\[2mm]&&\hspace*{3.2mm}
-8\,\HBS_1 \HBS_{2,1}\HBS_{2,1,1,1,2}
+48\,\HBS_1 \HBS_3\HBS_{3,1,1,1,1}
+8\,\HBS_1 \HBS_{2,1}\HBS_{3,1,1,1,1}\nonumber\\[2mm]&&\hspace*{3.2mm}
-12\Big(\HBS_1\HBS_2
-\HBS_3
-\HBS_{2,1}\Big)^2\HBS_{2,1,2}
+12\Big(\HBS_1\HBS_2
-\HBS_3
-\HBS_{2,1}\Big)^2\HBS_{3,1,1}
\Bigg)\,.\label{ResultRat}
\eeqa

So, collecting all terms we can obtain the general expression for $\cP_{12}^{\Wrap}$. The general result for the full anomalous dimension can be found in the ancillary files of the arXiv version of the paper and on the web-page: \href{http://thd.pnpi.spb.ru/~velizh/6loop/}{\texttt{http://thd.pnpi.spb.ru/\textasciitilde velizh/6loop/}}.

\section{Weak-coupling constraints}\label{sec:weak}

In this section we discuss the known weak-coupling constraints on the six-loop anomalous dimension of twist-two operators in $\cN=4$ SYM theory and verify that our result is consistent with them. We will use three classes of constraints, which are provided by the BFKL equation and by the generalised double-logarithmic equation at $\M=-2+\omega$ and at $\M=-r+\omega$, where $r=4,\,6,\,8,...$.

A splitting function $P(x)$, which is related to the anomalous dimension through a Mellin transformation
\beq
\gamma(\M)=\int^1_0 dx\, x^\M P(x)\,,
\eeq
contains some powers of $\ln x$ in each order of perturbative theory when $x$ tends to zero. This region of small $x$ is very interesting from the experimental point of view as there are some theoretical methods which allow one to sum all such large logarithms in all orders of perturbative theory. Mostly small-$x$ physics is related to the Balitsky-Fadin-Kuraev-Lipatov (BFKL) equation \cite{Lipatov:1976zz,Kuraev:1977fs,Balitsky:1978ic} and double-logarithmic equations~\cite{Kirschner:1982qf,Kirschner:1982xw,Kirschner:1983di}.

Since the splitting function is related to the anomalous dimension through a Mellin transformation, an expansion near $x=0$ corresponds to providing certain information about poles in the anomalous dimension analytically continued to negative values of $M$. A non-trivial pole structure should indeed be present  in the anomalous dimension since the harmonic sums~(\ref{vhs}), being a generalization of the $\Psi$-function, have poles at negative values of their argument. Hence the small-$x$ physics put a significant amount of constraints on the possible values of the conformal dimensions.

\subsection{BFKL equation} \label{sec:BFKL}

The relation between the anomalous dimension of twist-two operators and the Balitsky-Fadin-Kuraev-Lipatov (BFKL) equation \cite{Lipatov:1976zz,Kuraev:1977fs,Balitsky:1978ic} and its next-to-leading logarithm approximation (NLLA) generalisation \cite{Fadin:1998py,Kotikov:2000pm}  emerges upon analytic continuation of the function $\gamma(g,M)$ to complex values of $\M$. This is straightforward in the one-loop case since
\beq
\gamma_{2}(M) = 8\,g^2\,S_1 (M) = 8\,g^2\, \left(\Psi(M+1)-\Psi(1)\right)\,,
\eeq
where $\Psi(x)=\frac{d}{dx}\,\log \Gamma(x)$ is the digamma function.
At any loop order one expects singularities at all negative integer values of  $\M$. The first in this series of singular points,
\begin{equation}
\label{omega}
M=-1+\omega\,,
\end{equation}
corresponds to the so-called BFKL pomeron. In the above formula $\omega$ should be considered infinitesimally small. The BFKL equation relates $\gamma(g)$ and $\omega$ in the vicinity of the point $\M=-1+\omega$. It predicts that, if expanded in $g$, the $\ell$-loop anomalous dimension $\gamma_{2 \ell} (\omega)$ exhibits poles in $\omega$. Moreover, the residues and the order of the poles can be derived directly from the BFKL equation. The BFKL equation has been formulated up to the next-to-leading logarithm approximation (NLLA) and determines the leading and next-to-leading poles of $\gamma_{2\ell}(\omega)$.
The NLLA BFKL equation for twist-two operators in $\cN=4$ SYM theory in the dimensional reduction scheme can be written as follows~\cite{Fadin:1998py,Kotikov:2000pm}
\begin{equation}\label{BFKLgen}
\frac{\omega}{-4\,g^2} = \chi (\gamma )-g^2\,\delta (\gamma )\,,
\end{equation}
where
\begin{eqnarray}\label{gammanlo}
\chi (\gamma ) &=&
\Psi\left(-\frac{\gamma}{2}\right)+\Psi\left(1+\frac{\gamma}{2}\right)-
2\,\Psi\left(1\right)\, ,\\[4mm]
\delta (\gamma ) &=&4\,\chi ^{\,\prime \prime } (\gamma )
+6\,\z3+2\,\z2\,\chi (\gamma )+4\,\chi (\gamma )\,\chi ^{\,\prime} (\gamma )  \nonumber \\[2mm]
& & -\frac{\pi^3}{\sin \frac{\pi \gamma}{2}}- 4\,\Phi \left(-\frac{\gamma}{2}
\right) -4\,\Phi \left(1+\frac{\gamma}{2} \right)\,.
\end{eqnarray}
The function $\Phi (\gamma )$ is given by
\begin{eqnarray}
\Phi (\gamma ) =~\sum_{k=0}^{\infty }\frac{(-1)^{k}} {(k+\gamma)^2 }\biggl[\Psi
\left(k+\gamma +1\right)-\Psi (1)\biggr]. \label{9}
\end{eqnarray}
Upon using the expansion \eqref{dimension}, one easily determines the leading singularity structure.
Perturbatively expanding the anomalous dimension in the argument of the right-hand sided functions $\chi(\gamma)$ and $\delta(\gamma)$ in eq.~(\ref{BFKLgen}) one can find a relation between the anomalous dimension at $\M=-1+\omega$ and highest poles order by order, which can be written as:
\begin{eqnarray}
 \gamma&=&\left(2+0\,\omega+\Op(\omega^2)\right)
\left(\frac{-4\,g^2}{\omega}\right) -\left(0+0\,\omega
+\Op(\omega^2)\right)\,\left(\frac{-4\,g^2}{\omega}\right)^2
\nonumber
\\[-0.2mm]
&&
+\left(0+\,\z3\,\omega +\Op(\omega^2) \right)\,\left(\frac{-4\,g^2}{\omega}\right)^3
-\left(4\,\z3+\frac{5}{4}\,\z4\,\omega +\Op(\omega^2)\right)\,\left(\frac{-4\,g^2}{\omega}\right)^4 \nonumber \\[-0.2mm]
&&
-\left(0+\bigg(2\,\z2\,\z3+16\,\z5\bigg)\,\omega+\Op(\omega^2)\right)\left(\frac{-4g^2}{\omega}\right)^5 \nonumber\\[-0.2mm]
&&
-\left(\bigg(0\,\z2\,\z3+4\,\z5\bigg)+\bigg(3\,\zp{3}{2}-\frac{143}{48}\,\z6\bigg)\,\omega+\Op(\omega^2)\right)\left(\frac{-4g^2}{\omega}\right)^6\pm \ldots .
\label{BFKLPredictions}
\end{eqnarray}
The last line of the above equation gives the prediction for the analytic continuation of the six-loop anomalous dimension at $\M=-1+\omega$ and our full result satisfies these constraints\footnote{In \cite{Alfimov:2014bwa} it was shown that the leading pole computed from integrability should reproduce the BFKL prediction at any loop order.}.

\subsection{Generalised double-logarithmic equation at $\M=-2+\omega$} \label{sec:GenerDL2}

Another class of constraints on the anomalous dimension of twist-two operators follows from the double-logarithmic asymptotics of the scattering amplitudes.
The double-logarithmic asymptotics of the scattering amplitudes were studied in QED and QCD in the papers~\cite{Gorshkov:1966ht,Gorshkov:1966hu,Gorshkov:1966qd} and~\cite{Kirschner:1982qf,Kirschner:1982xw,Kirschner:1983di}
(see also {\texttt{arXiv}} version of ref.~\cite{Kotikov:2002ab}). It corresponds to summing the leading terms $(\alpha \ln^2s)^k$ in all orders of perturbation theory.
In combination with a Mellin transformation, the double-logarithmic asymptotics allow one to predict the singular part of the anomalous dimension near the point $\M=-2$.
For our purpose and in our notation the double-logarithmic equation has the following form:
\begin{equation}\label{DL}
\gamma\,(2\,\omega+\gamma)=-16\,g^2\,.
\end{equation}
The solution of this equation gives a prediction for the highest pole $(g^{2k}/\omega^{2k-1})$ in all orders of perturbative theory:
\begin{eqnarray}\label{dlevenp}
\gamma&=&-\omega+\omega\, \sqrt{1-\frac{16 g^2}{\omega^2}}
=
2\,\frac{(-4\, g^2)}{\omega}
-2\,\frac{(-4\, g^2)^2}{\omega^3}
+4\,\frac{(-4\, g^2)^3}{\omega^5}
-10\,\frac{(-4\, g^2)^4}{\omega^7}\nonumber \\
&&\hspace*{40mm}
+28\,\frac{(-4\, g^2)^5}{\omega^9}
-84\,\frac{(-4\, g^2)^6}{\omega^{11}}
+\ldots\, .\label{DLSolve}
\end{eqnarray}

The study of the analytic properties of the anomalous dimension of twist-two operators in $\cN=4$ SYM theory led to the suggestion about a simple generalisation of the double-logarithmic equation~\cite{Velizhanin:2011pb}\footnote{For the first time, such a generalisation was suggested by Lev N. Lipatov and Andrei Onishchenko in 2004,
but was not published. Then, it was improved by Lev N. Lipatov in ref.~\cite{Kotikov:2007cy}.}.
The main idea was that in eq.~(\ref{DL}) the corrections to the leading order equation will modify only the right-hand side and that such modification admits, besides an expansion in the coupling constant $g^2$, only the appearance of regular terms depending on $\omega$ (and, possibly, $\gamma$).
Substituting the results for the analytic continuation of the anomalous dimension of twist-two operators near $\M=-2+\omega$ into eq.~(\ref{DL}) we indeed find the following form of the generalised double-logarithmic equation~\cite{Velizhanin:2011pb}
\begin{equation}\label{DLgener}
\gamma\,(2\,\omega+\gamma)=\sum_{k=1}\sum_{m=0}{\mathfrak C}_m^k\,\omega^m\,g^{2k}\,.
\end{equation}
The coefficients ${\mathfrak C}_m^k$ can be found in the appendix of ref.~\cite{Velizhanin:2013vla}, but for the test of the six-loop anomalous dimension we need only the fact that the perturbative expansion of the left hand side of the above equation~(\ref{DLgener}) does not contain any poles near $\M=-2$, which is indeed correct for the result  obtained in this paper. We present the result for the analytic continuation of the six-loop anomalous dimension at $\M=-2+\omega$ in Appendix B, and one can see that there are about {\em{one hundred}} terms up to $g^{12}/w^2$, which can be checked by the generalised double-logarithmic equation~(\ref{DLgener}), so we have a very strong test for the correctness of our general result for the six-loop anomalous dimension.
Note also that with the generalised double-logarithmic equation~(\ref{DLgener}) we can control even the $\z9$ term in the six-loop anomalous dimension coming from eq.~(\ref{T12z9}) as its analytic continuation is proportional to $\z9/\omega^2$, what is impossible with the BFKL equation~(\ref{gammanlo}).

\subsection{Generalised double-logarithmic equation: $\M=-r+\omega,\ r=2,\ 4,\ 6,\ \ldots$} \label{sec:GenerDL2r}
In ref.~\cite{Velizhanin:2011pb} another generalisation of the double-logarithmic equation was found which holds true not only for $\M=-2+\omega$, but for all other negative even values $\M=-r+\omega\,,\ r=2,4,6,\ldots$\ . This generalisation is related to the analytic properties of the reciprocity-respecting anomalous dimension $\cP(\M)$,
which near $\M=-r+\omega\,,\ r=2,4,6,\ldots$ can be written as
\begin{equation}
{\mathcal P}_{\mathrm{DL}}(\omega,r) =2\,\sum _{k=1}\sum _{m=0}{\mathcal{D}}^k_m(r)\,\omega^m
\left(\frac{-4\,g^2}{\omega}\right)^k\,,\label{PDLr}
\end{equation}
where some coefficients ${\mathcal{D}}^k_m(r)$ can be found in ref.~\cite{Velizhanin:2011pb}. For the test of the six-loop anomalous dimension we need only the fact that according to eq.~(\ref{PDLr}) the six-loop reciprocity-respecting anomalous dimension $\cP_{12}(\M)$ does not contain any poles higher than $1/\omega^6$, which is indeed correct for the obtained result.

\subsection{Small $\M$ expansion}\label{sec:smallM}

The anomalous dimension for $\M=0$ is equal to zero while the expansion near $\M=0+\omega$ can be written as
\beqa
\gamma(0+\omega)= \sum_{k=1} \mathring{\gamma}_k\,\omega^k \,.
\eeqa
In ref.~\cite{Basso:2011rs} the expression for the first term in this expansion at arbitrary coupling was conjectured, and later it was confirmed in \cite{ Basso:2012ex, Gromov:2012eg}. The next term was obtained in ref.~\cite{Gromov:2014bva}.

The expansion near $\M=0+\omega$ can easily be obtained with the Mellin transformation of the six-loop splitting function $P(x)$ using
\beqa
\gamma(0+\omega)=\int^1_0\! dx\, x^{0+\omega} P(x)=\int^1_0\! dx\, P(x) \sum_{k=1} \frac{\ln^k\! x}{k!}\,\omega^k\,.
\eeqa
We have found  the six-loop splitting function $P(x)$ with the help of the \texttt{HARMPOL} package~\cite{Remiddi:1999ew} for \texttt{FORM}~\cite{Vermaseren:2000nd}, by expanding $P(x)$ in $x$ up to the order $x^{200}$ and then numerically integrating the obtained expressions with the help of \texttt{MATHEMATICA}, we computed the small spin expansion of the anomalous dimension. For the expansion up to $\omega^5$ the numerical results are
\beqa
\gamma(0+\omega)&=&
g^2\Big(
1.64493
\,\omega
-1.20206
\,\omega^2
+1.08232
\,\omega^3
-1.03693
\,\omega^4
+1.01734
\,\omega^5\Big)
\nonumber\\&&\hspace*{-18.23mm}
+g^4\Big(
-43.2925
\,\omega
+18.6224
\,\omega^2
-8.66878
\,\omega^3
+3.95542
\,\omega^4
-1.70134
\,\omega^5\Big)
\nonumber\\&&\hspace*{-18.23mm}
+g^6\Big(
227.767
\,\omega
+13.9824
\,\omega^2
-93.4614
\,\omega^3
+92.2139
\,\omega^4
-47.0089
\,\omega^5\Big)
\nonumber\\&&\hspace*{-18.23mm}
+g^8\Big(
-1315.67
\,\omega
-4460.78
\,\omega^2
+12751.98
\,\omega^3
-26031.74
\,\omega^4
+46274.86
\,\omega^5\Big)
\nonumber\\&&\hspace*{-18.23mm}
+g^{10}\Big(
8217.12
\,\omega
+97577.1
\,\omega^2
-290248
\,\omega^3
+618214
\,\omega^4
-1118763
\,\omega^5\Big)
\nonumber\\&&\hspace*{-18.23mm}
+g^{12}\Big(
-28323.3
\,\omega
-1528562
\,\omega^2
+4565909
\,\omega^3
-9705571
\,\omega^4
+17419037
\,\omega^5\Big)\,.
\eeqa
The coefficients of $\omega$ and $\omega^2$ coincide numerically correspondingly with~\cite{Basso:2011rs} and~\cite{Gromov:2014bva}. The coefficients for $\omega^3$ up to three loops are written in~\cite{Gromov:2014bva} and our results coincide numerically. Higher orders in $\omega$ is our prediction for the small-$M$ expansion. In pricniple, it would be also possible to derive these coefficients analytically if needed.

\section{Conclusion} \label{sec:discussion}

In this paper we have computed the full planar six-loop anomalous dimension of twist-two operators in $\cN=4$ SYM theory with arbitrary Lorentz spin $\M$.
The result is the sum of two parts, coming from ABA and from wrapping corrections, which were computed separately.
To find the most lengthy part, the rational contribution coming from ABA presented in Appendix~\ref{SpecialSums}, we computed its first 1024 values at fixed $\M$ which allowed us to fix the coefficients in the corresponding ansatz -- the linear combination of 1024 binomial harmonic sums that satisfy the maximal transcendentality principle.
To reconstruct the wrapping correction terms we used a {\texttt {MATHEMATICA}} implementation of the QSC approach~\cite{Marboe:2014gma} and  computed the six-loop anomalous dimension for 40 different fixed values of $\M$.
The corresponding ansatz for the rational part contains more than 300 binomial harmonic sums~(\ref{BinomialSums}).
As the rank of the system of linear equations is significantly smaller than its size we used a special method from number theory to find all coefficients in the ansatz.
With the help of the floating point realization~\cite{fplll} of the LLL-algorithm~\cite{Lenstra:1982} we obtained an LLL-reduced matrix which was formed from our linear system of Diophantine equations\footnote{We
expect that the physical solution of this system is given by integer numbers, hence the system can be identified as Diophantine equations. Moreover, we expect that the integers solving the equation are reasonably small in magnitude.}. The first row in the obtained matrix was the desired result, which is presented in eq.~(\ref{ResultRat}).

The parts which contain $\zeta_i$ are much simpler and can be reconstructed either from the direct solution of the system of linear equations or with the help of the LLL-algorithm. See Table~\ref{table:res} for an overview of where these terms are listed in this paper.  

All of the above concerns the computation of the reciprocity-respecting function $\reciP(M)$. The anomalous dimension is generated from $\reciP(M)$ and the lower-loop results with the help of eq.~(\ref{Pfunction}).
The expressions for the anomalous dimension are very lengthy, so we did not write them explicitly in the paper but they are available in the ancillary files of the arXiv version of the paper and on the web-page:
\href{http://thd.pnpi.spb.ru/~velizh/6loop/}{\texttt{http://thd.pnpi.spb.ru/\textasciitilde
velizh/6loop/}}.
\begin{table}[t]
\centering
\begin{tabular}{|c|c|c|c|c|c|c|c|c|c|c}
  \hline
                             &                   &       &       & &       &          &         &       &\\[-2mm]
         Contribution        & Rational          & $\z3$ & $\z5$ &$\zp{3}{2}$ & $\z7$ & $\z5\z3$ & $\zp{3}{3}$ & $\z9$ &Total\\[2mm]
  \hline
                             &                   &       &       & &       &          &         &       &\\[-2mm]
  $\gamma_{\rm{ABA}}$        &  (\ref{P12})     & (\ref{Pzt})   &(\ref{Pzf})    &       & (\ref{Pzs})    &         &        &      &(\ref{gammaABA})\\[1mm]
  $\reciP_{\rm{ABA}}$        &  (\ref{ABArat})   & (\ref{ABAz3}) &(\ref{ABAz5})  &       & (\ref{ABAz7})  &         &        &      &(\ref{PABAExp})\\[1mm]
                             &                   &       &       & &       &          &         &       &\\[-2mm]
  $\gamma_{\rm{Wrapping}}$   &  (\ref{gamma12P})   & (\ref{gamma12P})    &(\ref{gamma12P})    & (\ref{gamma12P})    & (\ref{gamma12P})     &  &        &      & (\ref{gamma12P})\\[1mm]
  $\reciP_{\rm{Wrapping}}$   &  (\ref{ResultRat})   & (\ref{Resultz3})& (\ref{Resultz5})     & (\ref{Resultz32})       & (\ref{Resultz7})& (\ref{T12z3z5})        & (\ref{T12z33})       & (\ref{T12z9})     &
(\ref{P12AnsatzZs})\\[2mm]
  \hline
\end{tabular}
\caption{Labels of equations with different contributions to our final
result.}
\label{table:res}
\end{table}

The obtained result was thoroughly tested against the constraints coming from the BFKL equation~(\ref{BFKLPredictions}) and the generalised double-logarithmic equations~(\ref{DLgener}), (\ref{PDLr}).
These equations provide more than {\it one hundred} constraints.
We found a complete agreement with these constraints, which confirms the correctness of the result.


\acknowledgments

The authors would like to thank A. Bednyakov, L.N. Lipatov, T. \L ukowski, A. Onishchenko, A. Ray, M. Staudacher, D. Stehl\'{e} for useful discussions.

The research of V.N. Velizhanin is supported by a Marie Curie International Incoming Fellowship within the 7th European Community Framework Programme, grant number PIIF-GA-2012-331484,
by DFG SFB 647 ``Raum -- Zeit -- Materie. Analytische und Geometrische Strukturen''and by RFBR grants 13-02-01246-a.
The research leading to these results has received funding from the People Programme (Marie Curie Actions) of the 7th European Community Framework Programme under REA Grant Agreement No 317089 (GATIS).
\newpage
\appendix
\section{Rational part of the six-loop reciprocity-respecting function $\cP_{12}^{\mathrm {ABA}}$} \label{SpecialSums}

We have found the following result for $\hat\cP_{12}(M)$ from eq.~(\ref{PABAExp}):
\beqa
\frac{{\hat\cP^{\textrm{rational}}_{12}}}{128}&=&
-180\,\HBS_{2,2,7}
+540\,\HBS_{3,1,7}
-592\,\HBS_{3,2,6}
+644\,\HBS_{4,1,6}
-684\,\HBS_{4,2,5}
+700\,\HBS_{5,1,5}\nonumber\\[2mm]&&\hspace*{-9mm}
-700\,\HBS_{5,2,4}
+636\,\HBS_{6,1,4}
-444\,\HBS_{6,2,3}
+176\,\HBS_{7,1,3}
-96\,\HBS_{7,2,2}
-228\,\HBS_{1,2,2,6}\nonumber\\[2mm]&&\hspace*{-9mm}
-4\,\HBS_{1,2,7,1}
+700\,\HBS_{1,3,1,6}
-728\,\HBS_{1,3,2,5}
+732\,\HBS_{1,4,1,5}
-728\,\HBS_{1,4,2,4}
+660\,\HBS_{1,5,1,4}\nonumber\\[2mm]&&\hspace*{-9mm}
-548\,\HBS_{1,5,2,3}
+344\,\HBS_{1,6,1,3}
-224\,\HBS_{1,6,2,2}
+52\,\HBS_{1,7,1,2}
-28\,\HBS_{1,7,2,1}
+4\,\HBS_{2,1,2,6}\nonumber\\[2mm]&&\hspace*{-9mm}
-4\,\HBS_{2,1,7,1}
-640\,\HBS_{2,2,1,6}
+988\,\HBS_{2,2,2,5}
+78\,\HBS_{2,2,3,4}
+78\,\HBS_{2,2,4,3}
+78\,\HBS_{2,2,5,2}\nonumber\\[2mm]&&\hspace*{-9mm}
-70\,\HBS_{2,2,6,1}
-864\,\HBS_{2,3,1,5}
+874\,\HBS_{2,3,2,4}
+2\,\HBS_{2,3,5,1}
-820\,\HBS_{2,4,1,4}
+698\,\HBS_{2,4,2,3}\nonumber\\[2mm]&&\hspace*{-9mm}
+2\,\HBS_{2,4,4,1}
-478\,\HBS_{2,5,1,3}
+326\,\HBS_{2,5,2,2}
+2\,\HBS_{2,5,3,1}
-106\,\HBS_{2,6,1,2}
+72\,\HBS_{2,6,2,1}\nonumber\\[2mm]&&\hspace*{-9mm}
-16\,\HBS_{2,7,1,1}
+592\,\HBS_{3,1,1,6}
-676\,\HBS_{3,1,2,5}
-242\,\HBS_{3,1,3,4}
-242\,\HBS_{3,1,4,3}
-242\,\HBS_{3,1,5,2}\nonumber\\[2mm]&&\hspace*{-9mm}
+282\,\HBS_{3,1,6,1}
-768\,\HBS_{3,2,1,5}
+1144\,\HBS_{3,2,2,4}
+260\,\HBS_{3,2,3,3}
+260\,\HBS_{3,2,4,2}
-288\,\HBS_{3,2,5,1}\nonumber\\[2mm]&&\hspace*{-9mm}
-290\,\HBS_{3,3,1,4}
+230\,\HBS_{3,3,2,3}
-138\,\HBS_{3,4,1,3}
+84\,\HBS_{3,4,2,2}
-14\,\HBS_{3,5,1,2}
+8\,\HBS_{3,5,2,1}\nonumber\\[2mm]&&\hspace*{-9mm}
+660\,\HBS_{4,1,1,5}
-746\,\HBS_{4,1,2,4}
-278\,\HBS_{4,1,3,3}
-278\,\HBS_{4,1,4,2}
+286\,\HBS_{4,1,5,1}
-762\,\HBS_{4,2,1,4}\nonumber\\[2mm]&&\hspace*{-9mm}
+1092\,\HBS_{4,2,2,3}
+296\,\HBS_{4,2,3,2}
-290\,\HBS_{4,2,4,1}
-138\,\HBS_{4,3,1,3}
+84\,\HBS_{4,3,2,2}
-14\,\HBS_{4,4,1,2}\nonumber\\[2mm]&&\hspace*{-9mm}
+8\,\HBS_{4,4,2,1}
+636\,\HBS_{5,1,1,4}
-726\,\HBS_{5,1,2,3}
-302\,\HBS_{5,1,3,2}
+276\,\HBS_{5,1,4,1}
-470\,\HBS_{5,2,1,3}\nonumber\\[2mm]&&\hspace*{-9mm}
+778\,\HBS_{5,2,2,2}
-274\,\HBS_{5,2,3,1}
-14\,\HBS_{5,3,1,2}
+8\,\HBS_{5,3,2,1}
+368\,\HBS_{6,1,1,3}
-492\,\HBS_{6,1,2,2}\nonumber\\[2mm]&&\hspace*{-9mm}
+240\,\HBS_{6,1,3,1}
-144\,\HBS_{6,2,1,2}
+16\,\HBS_{6,2,2,1}
+80\,\HBS_{7,1,1,2}
-32\,\HBS_{7,1,2,1}
-32\,\HBS_{7,2,1,1}\nonumber\\[2mm]&&\hspace*{-9mm}
-8\,\HBS_{1,1,1,2,6}
+8\,\HBS_{1,1,1,7,1}
-8\,\HBS_{1,1,2,1,6}
-236\,\HBS_{1,1,2,2,5}
+4\,\HBS_{1,1,2,3,4}
+4\,\HBS_{1,1,2,4,3}\nonumber\\[2mm]&&\hspace*{-9mm}
+4\,\HBS_{1,1,2,5,2}
-28\,\HBS_{1,1,2,6,1}
+788\,\HBS_{1,1,3,1,5}
-776\,\HBS_{1,1,3,2,4}
-4\,\HBS_{1,1,3,5,1}
+700\,\HBS_{1,1,4,1,4}\nonumber\\[2mm]&&\hspace*{-9mm}
-620\,\HBS_{1,1,4,2,3}
-4\,\HBS_{1,1,4,4,1}
+440\,\HBS_{1,1,5,1,3}
-324\,\HBS_{1,1,5,2,2}
-4\,\HBS_{1,1,5,3,1}
+128\,\HBS_{1,1,6,1,2}\nonumber\\[2mm]&&\hspace*{-9mm}
-96\,\HBS_{1,1,6,2,1}
+32\,\HBS_{1,1,7,1,1}
-8\,\HBS_{1,2,1,1,6}
+20\,\HBS_{1,2,1,2,5}
+4\,\HBS_{1,2,1,3,4}
+4\,\HBS_{1,2,1,4,3}\nonumber\\[2mm]&&\hspace*{-9mm}
+4\,\HBS_{1,2,1,5,2}
-24\,\HBS_{1,2,1,6,1}
-736\,\HBS_{1,2,2,1,5}
+1070\,\HBS_{1,2,2,2,4}
+82\,\HBS_{1,2,2,3,3}
+82\,\HBS_{1,2,2,4,2}\nonumber\\[2mm]&&\hspace*{-9mm}
-22\,\HBS_{1,2,2,5,1}
-870\,\HBS_{1,2,3,1,4}
+784\,\HBS_{1,2,3,2,3}
-2\,\HBS_{1,2,3,3,2}
+18\,\HBS_{1,2,3,4,1}
-600\,\HBS_{1,2,4,1,3}\nonumber\\[2mm]&&\hspace*{-9mm}
+454\,\HBS_{1,2,4,2,2}
+18\,\HBS_{1,2,4,3,1}
-214\,\HBS_{1,2,5,1,2}
+176\,\HBS_{1,2,5,2,1}
-76\,\HBS_{1,2,6,1,1}
+704\,\HBS_{1,3,1,1,5}\nonumber\\[2mm]&&\hspace*{-9mm}
-794\,\HBS_{1,3,1,2,4}
-302\,\HBS_{1,3,1,3,3}
-302\,\HBS_{1,3,1,4,2}
+306\,\HBS_{1,3,1,5,1}
-794\,\HBS_{1,3,2,1,4}
+1174\,\HBS_{1,3,2,2,3}\nonumber\\[2mm]&&\hspace*{-9mm}
+312\,\HBS_{1,3,2,3,2}
-304\,\HBS_{1,3,2,4,1}
-198\,\HBS_{1,3,3,1,3}
+140\,\HBS_{1,3,3,2,2}
+2\,\HBS_{1,3,3,3,1}
-54\,\HBS_{1,3,4,1,2}\nonumber\\[2mm]&&\hspace*{-9mm}
+38\,\HBS_{1,3,4,2,1}
-12\,\HBS_{1,3,5,1,1}
+664\,\HBS_{1,4,1,1,4}
-756\,\HBS_{1,4,1,2,3}
-310\,\HBS_{1,4,1,3,2}
+292\,\HBS_{1,4,1,4,1}\nonumber
\eeqa
\beqa
&&\hspace*{2mm}
-576\,\HBS_{1,4,2,1,3}
+904\,\HBS_{1,4,2,2,2}
-286\,\HBS_{1,4,2,3,1}
-54\,\HBS_{1,4,3,1,2}
+38\,\HBS_{1,4,3,2,1}
-12\,\HBS_{1,4,4,1,1}\nonumber\\[2mm]&&\hspace*{2mm}
+456\,\HBS_{1,5,1,1,3}
-568\,\HBS_{1,5,1,2,2}
+258\,\HBS_{1,5,1,3,1}
-244\,\HBS_{1,5,2,1,2}
+110\,\HBS_{1,5,2,2,1}
-12\,\HBS_{1,5,3,1,1}\nonumber\\[2mm]&&\hspace*{2mm}
+172\,\HBS_{1,6,1,1,2}
-92\,\HBS_{1,6,1,2,1}
-120\,\HBS_{1,6,2,1,1}
+56\,\HBS_{1,7,1,1,1}
-8\,\HBS_{2,1,1,1,6}
+20\,\HBS_{2,1,1,2,5}\nonumber\\[2mm]&&\hspace*{2mm}
+4\,\HBS_{2,1,1,3,4}
+4\,\HBS_{2,1,1,4,3}
+4\,\HBS_{2,1,1,5,2}
-24\,\HBS_{2,1,1,6,1}
+16\,\HBS_{2,1,2,1,5}
+68\,\HBS_{2,1,2,2,4}\nonumber\\[2mm]&&\hspace*{2mm}
-14\,\HBS_{2,1,2,3,3}
-14\,\HBS_{2,1,2,4,2}
+56\,\HBS_{2,1,2,5,1}
-340\,\HBS_{2,1,3,1,4}
+316\,\HBS_{2,1,3,2,3}
-2\,\HBS_{2,1,3,3,2}\nonumber\\[2mm]&&\hspace*{2mm}
+16\,\HBS_{2,1,3,4,1}
-260\,\HBS_{2,1,4,1,3}
+212\,\HBS_{2,1,4,2,2}
+16\,\HBS_{2,1,4,3,1}
-122\,\HBS_{2,1,5,1,2}
+112\,\HBS_{2,1,5,2,1}\nonumber\\[2mm]&&\hspace*{2mm}
-60\,\HBS_{2,1,6,1,1}
-640\,\HBS_{2,2,1,1,5}
+704\,\HBS_{2,2,1,2,4}
+264\,\HBS_{2,2,1,3,3}
+264\,\HBS_{2,2,1,4,2}
-240\,\HBS_{2,2,1,5,1}\nonumber\\[2mm]&&\hspace*{2mm}
+1226\,\HBS_{2,2,2,1,4}
-1816\,\HBS_{2,2,2,2,3}
-381\,\HBS_{2,2,2,3,2}
+245\,\HBS_{2,2,2,4,1}
+638\,\HBS_{2,2,3,1,3}
-540\,\HBS_{2,2,3,2,2}\nonumber\\[2mm]&&\hspace*{2mm}
-18\,\HBS_{2,2,3,3,1}
+289\,\HBS_{2,2,4,1,2}
-179\,\HBS_{2,2,4,2,1}
+56\,\HBS_{2,2,5,1,1}
-810\,\HBS_{2,3,1,1,4}
+918\,\HBS_{2,3,1,2,3}\nonumber\\[2mm]&&\hspace*{2mm}
+367\,\HBS_{2,3,1,3,2}
-339\,\HBS_{2,3,1,4,1}
+742\,\HBS_{2,3,2,1,3}
-1136\,\HBS_{2,3,2,2,2}
+322\,\HBS_{2,3,2,3,1}
+93\,\HBS_{2,3,3,1,2}\nonumber\\[2mm]&&\hspace*{2mm}
-77\,\HBS_{2,3,3,2,1}
+36\,\HBS_{2,3,4,1,1}
-600\,\HBS_{2,4,1,1,3}
+730\,\HBS_{2,4,1,2,2}
-312\,\HBS_{2,4,1,3,1}
+358\,\HBS_{2,4,2,1,2}\nonumber\\[2mm]&&\hspace*{2mm}
-200\,\HBS_{2,4,2,2,1}
+36\,\HBS_{2,4,3,1,1}
-258\,\HBS_{2,5,1,1,2}
+136\,\HBS_{2,5,1,2,1}
+206\,\HBS_{2,5,2,1,1}
-116\,\HBS_{2,6,1,1,1}\nonumber\\[2mm]&&\hspace*{2mm}
+620\,\HBS_{3,1,1,1,5}
-700\,\HBS_{3,1,1,2,4}
-260\,\HBS_{3,1,1,3,3}
-260\,\HBS_{3,1,1,4,2}
+280\,\HBS_{3,1,1,5,1}
-266\,\HBS_{3,1,2,1,4}\nonumber\\[2mm]&&\hspace*{2mm}
+602\,\HBS_{3,1,2,2,3}
+293\,\HBS_{3,1,2,3,2}
-307\,\HBS_{3,1,2,4,1}
+50\,\HBS_{3,1,3,1,3}
+84\,\HBS_{3,1,3,2,2}
-110\,\HBS_{3,1,3,3,1}\nonumber\\[2mm]&&\hspace*{2mm}
+65\,\HBS_{3,1,4,1,2}
-371\,\HBS_{3,1,4,2,1}
+348\,\HBS_{3,1,5,1,1}
-738\,\HBS_{3,2,1,1,4}
+832\,\HBS_{3,2,1,2,3}
+329\,\HBS_{3,2,1,3,2}\nonumber\\[2mm]&&\hspace*{2mm}
-309\,\HBS_{3,2,1,4,1}
+550\,\HBS_{3,2,2,1,3}
-1126\,\HBS_{3,2,2,2,2}
+456\,\HBS_{3,2,2,3,1}
-31\,\HBS_{3,2,3,1,2}
+347\,\HBS_{3,2,3,2,1}\nonumber\\[2mm]&&\hspace*{2mm}
-342\,\HBS_{3,2,4,1,1}
-198\,\HBS_{3,3,1,1,3}
+248\,\HBS_{3,3,1,2,2}
-108\,\HBS_{3,3,1,3,1}
+102\,\HBS_{3,3,2,1,2}
-48\,\HBS_{3,3,2,2,1}\nonumber\\[2mm]&&\hspace*{2mm}
+4\,\HBS_{3,3,3,1,1}
-68\,\HBS_{3,4,1,1,2}
+36\,\HBS_{3,4,1,2,1}
+44\,\HBS_{3,4,2,1,1}
-16\,\HBS_{3,5,1,1,1}
+616\,\HBS_{4,1,1,1,4}\nonumber\\[2mm]&&\hspace*{2mm}
-700\,\HBS_{4,1,1,2,3}
-284\,\HBS_{4,1,1,3,2}
+272\,\HBS_{4,1,1,4,1}
-232\,\HBS_{4,1,2,1,3}
+594\,\HBS_{4,1,2,2,2}
-304\,\HBS_{4,1,2,3,1}\nonumber\\[2mm]&&\hspace*{2mm}
+60\,\HBS_{4,1,3,1,2}
-362\,\HBS_{4,1,3,2,1}
+344\,\HBS_{4,1,4,1,1}
-564\,\HBS_{4,2,1,1,3}
+684\,\HBS_{4,2,1,2,2}
-288\,\HBS_{4,2,1,3,1}\nonumber\\[2mm]&&\hspace*{2mm}
+280\,\HBS_{4,2,2,1,2}
+188\,\HBS_{4,2,2,2,1}
-336\,\HBS_{4,2,3,1,1}
-68\,\HBS_{4,3,1,1,2}
+36\,\HBS_{4,3,1,2,1}
+44\,\HBS_{4,3,2,1,1}\nonumber\\[2mm]&&\hspace*{2mm}
-16\,\HBS_{4,4,1,1,1}
+452\,\HBS_{5,1,1,1,3}
-556\,\HBS_{5,1,1,2,2}
+242\,\HBS_{5,1,1,3,1}
-132\,\HBS_{5,1,2,1,2}
-192\,\HBS_{5,1,2,2,1}\nonumber\\[2mm]&&\hspace*{2mm}
+330\,\HBS_{5,1,3,1,1}
-262\,\HBS_{5,2,1,1,2}
+140\,\HBS_{5,2,1,2,1}
+98\,\HBS_{5,2,2,1,1}
-16\,\HBS_{5,3,1,1,1}
+188\,\HBS_{6,1,1,1,2}\nonumber\\[2mm]&&\hspace*{2mm}
-96\,\HBS_{6,1,1,2,1}
-96\,\HBS_{6,1,2,1,1}
-128\,\HBS_{6,2,1,1,1}
+64\,\HBS_{7,1,1,1,1}
+16\,\HBS_{1,1,1,1,1,6}
-24\,\HBS_{1,1,1,1,2,5}\nonumber\\[2mm]&&\hspace*{2mm}
-8\,\HBS_{1,1,1,1,3,4}
-8\,\HBS_{1,1,1,1,4,3}
-8\,\HBS_{1,1,1,1,5,2}
+32\,\HBS_{1,1,1,1,6,1}
-16\,\HBS_{1,1,1,2,1,5}
-200\,\HBS_{1,1,1,2,2,4}\nonumber\\[2mm]&&\hspace*{2mm}
+20\,\HBS_{1,1,1,2,3,3}
+20\,\HBS_{1,1,1,2,4,2}
-68\,\HBS_{1,1,1,2,5,1}
+744\,\HBS_{1,1,1,3,1,4}
-668\,\HBS_{1,1,1,3,2,3}
+4\,\HBS_{1,1,1,3,3,2}\nonumber\\[2mm]&&\hspace*{2mm}
-24\,\HBS_{1,1,1,3,4,1}
+508\,\HBS_{1,1,1,4,1,3}
-392\,\HBS_{1,1,1,4,2,2}
-24\,\HBS_{1,1,1,4,3,1}
+192\,\HBS_{1,1,1,5,1,2}\nonumber
\eeqa
\beqa
&&\hspace*{2mm}
-176\,\HBS_{1,1,1,5,2,1}
+80\,\HBS_{1,1,1,6,1,1}
-16\,\HBS_{1,1,2,1,1,5}
+28\,\HBS_{1,1,2,1,2,4}
+16\,\HBS_{1,1,2,1,3,3}\nonumber\\[2mm]&&\hspace*{2mm}
+16\,\HBS_{1,1,2,1,4,2}
-44\,\HBS_{1,1,2,1,5,1}
-740\,\HBS_{1,1,2,2,1,4}
+974\,\HBS_{1,1,2,2,2,3}
+64\,\HBS_{1,1,2,2,3,2}\nonumber\\[2mm]&&\hspace*{2mm}
+38\,\HBS_{1,1,2,2,4,1}
-678\,\HBS_{1,1,2,3,1,3}
+534\,\HBS_{1,1,2,3,2,2}
+50\,\HBS_{1,1,2,3,3,1}
-308\,\HBS_{1,1,2,4,1,2}\nonumber\\[2mm]&&\hspace*{2mm}
+310\,\HBS_{1,1,2,4,2,1}
-164\,\HBS_{1,1,2,5,1,1}
+712\,\HBS_{1,1,3,1,1,4}
-810\,\HBS_{1,1,3,1,2,3}
-328\,\HBS_{1,1,3,1,3,2}\nonumber\\[2mm]&&\hspace*{2mm}
+310\,\HBS_{1,1,3,1,4,1}
-654\,\HBS_{1,1,3,2,1,3}
+1006\,\HBS_{1,1,3,2,2,2}
-282\,\HBS_{1,1,3,2,3,1}
-96\,\HBS_{1,1,3,3,1,2}\nonumber\\[2mm]&&\hspace*{2mm}
+90\,\HBS_{1,1,3,3,2,1}
-48\,\HBS_{1,1,3,4,1,1}
+520\,\HBS_{1,1,4,1,1,3}
-628\,\HBS_{1,1,4,1,2,2}
+276\,\HBS_{1,1,4,1,3,1}\nonumber\\[2mm]&&\hspace*{2mm}
-332\,\HBS_{1,1,4,2,1,2}
+200\,\HBS_{1,1,4,2,2,1}
-48\,\HBS_{1,1,4,3,1,1}
+244\,\HBS_{1,1,5,1,1,2}
-128\,\HBS_{1,1,5,1,2,1}\nonumber\\[2mm]&&\hspace*{2mm}
-220\,\HBS_{1,1,5,2,1,1}
+136\,\HBS_{1,1,6,1,1,1}
-16\,\HBS_{1,2,1,1,1,5}
+28\,\HBS_{1,2,1,1,2,4}
+16\,\HBS_{1,2,1,1,3,3}\nonumber\\[2mm]&&\hspace*{2mm}
+16\,\HBS_{1,2,1,1,4,2}
-44\,\HBS_{1,2,1,1,5,1}
+12\,\HBS_{1,2,1,2,1,4}
+36\,\HBS_{1,2,1,2,2,3}
-30\,\HBS_{1,2,1,2,3,2}\nonumber\\[2mm]&&\hspace*{2mm}
+80\,\HBS_{1,2,1,2,4,1}
-276\,\HBS_{1,2,1,3,1,3}
+220\,\HBS_{1,2,1,3,2,2}
+34\,\HBS_{1,2,1,3,3,1}
-148\,\HBS_{1,2,1,4,1,2}\nonumber\\[2mm]&&\hspace*{2mm}
+172\,\HBS_{1,2,1,4,2,1}
-100\,\HBS_{1,2,1,5,1,1}
-664\,\HBS_{1,2,2,1,1,4}
+742\,\HBS_{1,2,2,1,2,3}
+294\,\HBS_{1,2,2,1,3,2}\nonumber\\[2mm]&&\hspace*{2mm}
-244\,\HBS_{1,2,2,1,4,1}
+1078\,\HBS_{1,2,2,2,1,3}
-1582\,\HBS_{1,2,2,2,2,2}
+178\,\HBS_{1,2,2,2,3,1}
+398\,\HBS_{1,2,2,3,1,2}\nonumber\\[2mm]&&\hspace*{2mm}
-362\,\HBS_{1,2,2,3,2,1}
+186\,\HBS_{1,2,2,4,1,1}
-682\,\HBS_{1,2,3,1,1,3}
+804\,\HBS_{1,2,3,1,2,2}
-324\,\HBS_{1,2,3,1,3,1}\nonumber\\[2mm]&&\hspace*{2mm}
+474\,\HBS_{1,2,3,2,1,2}
-336\,\HBS_{1,2,3,2,2,1}
+94\,\HBS_{1,2,3,3,1,1}
-356\,\HBS_{1,2,4,1,1,2}
+188\,\HBS_{1,2,4,1,2,1}\nonumber\\[2mm]&&\hspace*{2mm}
+346\,\HBS_{1,2,4,2,1,1}
-232\,\HBS_{1,2,5,1,1,1}
+632\,\HBS_{1,3,1,1,1,4}
-724\,\HBS_{1,3,1,1,2,3}
-300\,\HBS_{1,3,1,1,3,2}\nonumber\\[2mm]&&\hspace*{2mm}
+290\,\HBS_{1,3,1,1,4,1}
-232\,\HBS_{1,3,1,2,1,3}
+616\,\HBS_{1,3,1,2,2,2}
-320\,\HBS_{1,3,1,2,3,1}
+60\,\HBS_{1,3,1,3,1,2}\nonumber\\[2mm]&&\hspace*{2mm}
-380\,\HBS_{1,3,1,3,2,1}
+362\,\HBS_{1,3,1,4,1,1}
-610\,\HBS_{1,3,2,1,1,3}
+728\,\HBS_{1,3,2,1,2,2}
-302\,\HBS_{1,3,2,1,3,1}\nonumber\\[2mm]&&\hspace*{2mm}
+346\,\HBS_{1,3,2,2,1,2}
+134\,\HBS_{1,3,2,2,2,1}
-332\,\HBS_{1,3,2,3,1,1}
-112\,\HBS_{1,3,3,1,1,2}
+64\,\HBS_{1,3,3,1,2,1}\nonumber\\[2mm]&&\hspace*{2mm}
+100\,\HBS_{1,3,3,2,1,1}
-60\,\HBS_{1,3,4,1,1,1}
+496\,\HBS_{1,4,1,1,1,3}
-598\,\HBS_{1,4,1,1,2,2}
+260\,\HBS_{1,4,1,1,3,1}\nonumber\\[2mm]&&\hspace*{2mm}
-152\,\HBS_{1,4,1,2,1,2}
-200\,\HBS_{1,4,1,2,2,1}
+348\,\HBS_{1,4,1,3,1,1}
-342\,\HBS_{1,4,2,1,1,2}
+188\,\HBS_{1,4,2,1,2,1}\nonumber\\[2mm]&&\hspace*{2mm}
+196\,\HBS_{1,4,2,2,1,1}
-60\,\HBS_{1,4,3,1,1,1}
+260\,\HBS_{1,5,1,1,1,2}
-136\,\HBS_{1,5,1,1,2,1}
-136\,\HBS_{1,5,1,2,1,1}\nonumber\\[2mm]&&\hspace*{2mm}
-244\,\HBS_{1,5,2,1,1,1}
+160\,\HBS_{1,6,1,1,1,1}
-16\,\HBS_{2,1,1,1,1,5}
+28\,\HBS_{2,1,1,1,2,4}
+16\,\HBS_{2,1,1,1,3,3}\nonumber\\[2mm]&&\hspace*{2mm}
+16\,\HBS_{2,1,1,1,4,2}
-44\,\HBS_{2,1,1,1,5,1}
+12\,\HBS_{2,1,1,2,1,4}
+36\,\HBS_{2,1,1,2,2,3}
-30\,\HBS_{2,1,1,2,3,2}\nonumber\\[2mm]&&\hspace*{2mm}
+80\,\HBS_{2,1,1,2,4,1}
-276\,\HBS_{2,1,1,3,1,3}
+220\,\HBS_{2,1,1,3,2,2}
+34\,\HBS_{2,1,1,3,3,1}
-148\,\HBS_{2,1,1,4,1,2}\nonumber\\[2mm]&&\hspace*{2mm}
+172\,\HBS_{2,1,1,4,2,1}
-100\,\HBS_{2,1,1,5,1,1}
+12\,\HBS_{2,1,2,1,1,4}
-26\,\HBS_{2,1,2,1,2,3}
-18\,\HBS_{2,1,2,1,3,2}\nonumber\\[2mm]&&\hspace*{2mm}
+40\,\HBS_{2,1,2,1,4,1}
+284\,\HBS_{2,1,2,2,1,3}
-298\,\HBS_{2,1,2,2,2,2}
-108\,\HBS_{2,1,2,2,3,1}
+222\,\HBS_{2,1,2,3,1,2}\nonumber\\[2mm]&&\hspace*{2mm}
-280\,\HBS_{2,1,2,3,2,1}
+180\,\HBS_{2,1,2,4,1,1}
-280\,\HBS_{2,1,3,1,1,3}
+322\,\HBS_{2,1,3,1,2,2}
-120\,\HBS_{2,1,3,1,3,1}\nonumber
\eeqa
\beqa
&&\hspace*{2mm}
+218\,\HBS_{2,1,3,2,1,2}
-184\,\HBS_{2,1,3,2,2,1}
+62\,\HBS_{2,1,3,3,1,1}
-166\,\HBS_{2,1,4,1,1,2}
+88\,\HBS_{2,1,4,1,2,1}\nonumber\\[2mm]&&\hspace*{2mm}
+184\,\HBS_{2,1,4,2,1,1}
-132\,\HBS_{2,1,5,1,1,1}
-592\,\HBS_{2,2,1,1,1,4}
+656\,\HBS_{2,2,1,1,2,3}
+256\,\HBS_{2,2,1,1,3,2}\nonumber\\[2mm]&&\hspace*{2mm}
-214\,\HBS_{2,2,1,1,4,1}
+216\,\HBS_{2,2,1,2,1,3}
-550\,\HBS_{2,2,1,2,2,2}
+198\,\HBS_{2,2,1,2,3,1}
+78\,\HBS_{2,2,1,3,1,2}\nonumber\\[2mm]&&\hspace*{2mm}
+164\,\HBS_{2,2,1,3,2,1}
-212\,\HBS_{2,2,1,4,1,1}
+990\,\HBS_{2,2,2,1,1,3}
-1131\,\HBS_{2,2,2,1,2,2}
+389\,\HBS_{2,2,2,1,3,1}\nonumber\\[2mm]&&\hspace*{2mm}
-892\,\HBS_{2,2,2,2,1,2}
+604\,\HBS_{2,2,2,2,2,1}
+36\,\HBS_{2,2,2,3,1,1}
+415\,\HBS_{2,2,3,1,1,2}
-187\,\HBS_{2,2,3,1,2,1}\nonumber\\[2mm]&&\hspace*{2mm}
-348\,\HBS_{2,2,3,2,1,1}
+220\,\HBS_{2,2,4,1,1,1}
-626\,\HBS_{2,3,1,1,1,3}
+747\,\HBS_{2,3,1,1,2,2}
-313\,\HBS_{2,3,1,1,3,1}\nonumber\\[2mm]&&\hspace*{2mm}
+196\,\HBS_{2,3,1,2,1,2}
+204\,\HBS_{2,3,1,2,2,1}
-386\,\HBS_{2,3,1,3,1,1}
+461\,\HBS_{2,3,2,1,1,2}
-257\,\HBS_{2,3,2,1,2,1}\nonumber\\[2mm]&&\hspace*{2mm}
-318\,\HBS_{2,3,2,2,1,1}
+100\,\HBS_{2,3,3,1,1,1}
-356\,\HBS_{2,4,1,1,1,2}
+190\,\HBS_{2,4,1,1,2,1}
+190\,\HBS_{2,4,1,2,1,1}\nonumber\\[2mm]&&\hspace*{2mm}
+360\,\HBS_{2,4,2,1,1,1}
-248\,\HBS_{2,5,1,1,1,1}
+592\,\HBS_{3,1,1,1,1,4}
-668\,\HBS_{3,1,1,1,2,3}
-264\,\HBS_{3,1,1,1,3,2}\nonumber\\[2mm]&&\hspace*{2mm}
+260\,\HBS_{3,1,1,1,4,1}
-228\,\HBS_{3,1,1,2,1,3}
+560\,\HBS_{3,1,1,2,2,2}
-292\,\HBS_{3,1,1,2,3,1}
+66\,\HBS_{3,1,1,3,1,2}\nonumber\\[2mm]&&\hspace*{2mm}
-364\,\HBS_{3,1,1,3,2,1}
+342\,\HBS_{3,1,1,4,1,1}
-228\,\HBS_{3,1,2,1,1,3}
+259\,\HBS_{3,1,2,1,2,2}
-95\,\HBS_{3,1,2,1,3,1}\nonumber\\[2mm]&&\hspace*{2mm}
+42\,\HBS_{3,1,2,2,1,2}
+426\,\HBS_{3,1,2,2,2,1}
-368\,\HBS_{3,1,2,3,1,1}
+59\,\HBS_{3,1,3,1,1,2}
-147\,\HBS_{3,1,3,1,2,1}\nonumber\\[2mm]&&\hspace*{2mm}
-408\,\HBS_{3,1,3,2,1,1}
+388\,\HBS_{3,1,4,1,1,1}
-594\,\HBS_{3,2,1,1,1,3}
+695\,\HBS_{3,2,1,1,2,2}
-275\,\HBS_{3,2,1,1,3,1}\nonumber\\[2mm]&&\hspace*{2mm}
+192\,\HBS_{3,2,1,2,1,2}
+172\,\HBS_{3,2,1,2,2,1}
-362\,\HBS_{3,2,1,3,1,1}
+353\,\HBS_{3,2,2,1,1,2}
-173\,\HBS_{3,2,2,1,2,1}\nonumber\\[2mm]&&\hspace*{2mm}
+158\,\HBS_{3,2,2,2,1,1}
-346\,\HBS_{3,2,3,1,1,1}
-114\,\HBS_{3,3,1,1,1,2}
+66\,\HBS_{3,3,1,1,2,1}
+66\,\HBS_{3,3,1,2,1,1}\nonumber\\[2mm]&&\hspace*{2mm}
+106\,\HBS_{3,3,2,1,1,1}
-64\,\HBS_{3,4,1,1,1,1}
+480\,\HBS_{4,1,1,1,1,3}
-568\,\HBS_{4,1,1,1,2,2}
+236\,\HBS_{4,1,1,1,3,1}\nonumber\\[2mm]&&\hspace*{2mm}
-152\,\HBS_{4,1,1,2,1,2}
-178\,\HBS_{4,1,1,2,2,1}
+330\,\HBS_{4,1,1,3,1,1}
-152\,\HBS_{4,1,2,1,1,2}
+86\,\HBS_{4,1,2,1,2,1}\nonumber\\[2mm]&&\hspace*{2mm}
-236\,\HBS_{4,1,2,2,1,1}
+366\,\HBS_{4,1,3,1,1,1}
-346\,\HBS_{4,2,1,1,1,2}
+190\,\HBS_{4,2,1,1,2,1}
+190\,\HBS_{4,2,1,2,1,1}\nonumber\\[2mm]&&\hspace*{2mm}
+198\,\HBS_{4,2,2,1,1,1}
-64\,\HBS_{4,3,1,1,1,1}
+264\,\HBS_{5,1,1,1,1,2}
-136\,\HBS_{5,1,1,1,2,1}
-136\,\HBS_{5,1,1,2,1,1}\nonumber\\[2mm]&&\hspace*{2mm}
-136\,\HBS_{5,1,2,1,1,1}
-248\,\HBS_{5,2,1,1,1,1}
+160\,\HBS_{6,1,1,1,1,1}
-16\,\HBS_{1,1,1,1,1,2,4}
-16\,\HBS_{1,1,1,1,1,3,3}\nonumber\\[2mm]&&\hspace*{2mm}
-16\,\HBS_{1,1,1,1,1,4,2}
+40\,\HBS_{1,1,1,1,1,5,1}
-140\,\HBS_{1,1,1,1,2,2,3}
+28\,\HBS_{1,1,1,1,2,3,2}
-72\,\HBS_{1,1,1,1,2,4,1}\nonumber\\[2mm]&&\hspace*{2mm}
+540\,\HBS_{1,1,1,1,3,1,3}
-420\,\HBS_{1,1,1,1,3,2,2}
-32\,\HBS_{1,1,1,1,3,3,1}
+232\,\HBS_{1,1,1,1,4,1,2}
-248\,\HBS_{1,1,1,1,4,2,1}\nonumber\\[2mm]&&\hspace*{2mm}
+120\,\HBS_{1,1,1,1,5,1,1}
+12\,\HBS_{1,1,1,2,1,2,3}
+12\,\HBS_{1,1,1,2,1,3,2}
-28\,\HBS_{1,1,1,2,1,4,1}
-608\,\HBS_{1,1,1,2,2,1,3}\nonumber\\[2mm]&&\hspace*{2mm}
+730\,\HBS_{1,1,1,2,2,2,2}
+44\,\HBS_{1,1,1,2,2,3,1}
-374\,\HBS_{1,1,1,2,3,1,2}
+436\,\HBS_{1,1,1,2,3,2,1}
-240\,\HBS_{1,1,1,2,4,1,1}\nonumber\\[2mm]&&\hspace*{2mm}
+576\,\HBS_{1,1,1,3,1,1,3}
-682\,\HBS_{1,1,1,3,1,2,2}
+288\,\HBS_{1,1,1,3,1,3,1}
-406\,\HBS_{1,1,1,3,2,1,2}
+284\,\HBS_{1,1,1,3,2,2,1}\nonumber\\[2mm]&&\hspace*{2mm}
-96\,\HBS_{1,1,1,3,3,1,1}
+300\,\HBS_{1,1,1,4,1,1,2}
-148\,\HBS_{1,1,1,4,1,2,1}
-300\,\HBS_{1,1,1,4,2,1,1}
+200\,\HBS_{1,1,1,5,1,1,1}\nonumber\\[2mm]&&\hspace*{2mm}
+12\,\HBS_{1,1,2,1,1,2,3}
+12\,\HBS_{1,1,2,1,1,3,2}
-28\,\HBS_{1,1,2,1,1,4,1}
+18\,\HBS_{1,1,2,1,2,2,2}
+60\,\HBS_{1,1,2,1,2,3,1}\nonumber
\eeqa
\beqa
&&\hspace*{2mm}
-162\,\HBS_{1,1,2,1,3,1,2}
+216\,\HBS_{1,1,2,1,3,2,1}
-124\,\HBS_{1,1,2,1,4,1,1}
-584\,\HBS_{1,1,2,2,1,1,3}
+680\,\HBS_{1,1,2,2,1,2,2}\nonumber\\[2mm]&&\hspace*{2mm}
-276\,\HBS_{1,1,2,2,1,3,1}
+762\,\HBS_{1,1,2,2,2,1,2}
-848\,\HBS_{1,1,2,2,2,2,1}
+298\,\HBS_{1,1,2,2,3,1,1}
-432\,\HBS_{1,1,2,3,1,1,2}\nonumber\\[2mm]&&\hspace*{2mm}
+222\,\HBS_{1,1,2,3,1,2,1}
+482\,\HBS_{1,1,2,3,2,1,1}
-348\,\HBS_{1,1,2,4,1,1,1}
+528\,\HBS_{1,1,3,1,1,1,3}
-638\,\HBS_{1,1,3,1,1,2,2}\nonumber\\[2mm]&&\hspace*{2mm}
+284\,\HBS_{1,1,3,1,1,3,1}
-156\,\HBS_{1,1,3,1,2,1,2}
-216\,\HBS_{1,1,3,1,2,2,1}
+358\,\HBS_{1,1,3,1,3,1,1}
-392\,\HBS_{1,1,3,2,1,1,2}\nonumber\\[2mm]&&\hspace*{2mm}
+206\,\HBS_{1,1,3,2,1,2,1}
+266\,\HBS_{1,1,3,2,2,1,1}
-104\,\HBS_{1,1,3,3,1,1,1}
+304\,\HBS_{1,1,4,1,1,1,2}
-156\,\HBS_{1,1,4,1,1,2,1}\nonumber\\[2mm]&&\hspace*{2mm}
-156\,\HBS_{1,1,4,1,2,1,1}
-328\,\HBS_{1,1,4,2,1,1,1}
+240\,\HBS_{1,1,5,1,1,1,1}
+12\,\HBS_{1,2,1,1,1,2,3}
+12\,\HBS_{1,2,1,1,1,3,2}\nonumber\\[2mm]&&\hspace*{2mm}
-28\,\HBS_{1,2,1,1,1,4,1}
+18\,\HBS_{1,2,1,1,2,2,2}
+60\,\HBS_{1,2,1,1,2,3,1}
-162\,\HBS_{1,2,1,1,3,1,2}
+216\,\HBS_{1,2,1,1,3,2,1}\nonumber\\[2mm]&&\hspace*{2mm}
-124\,\HBS_{1,2,1,1,4,1,1}
-8\,\HBS_{1,2,1,2,1,2,2}
+14\,\HBS_{1,2,1,2,1,3,1}
+202\,\HBS_{1,2,1,2,2,1,2}
-390\,\HBS_{1,2,1,2,2,2,1}\nonumber\\[2mm]&&\hspace*{2mm}
+208\,\HBS_{1,2,1,2,3,1,1}
-188\,\HBS_{1,2,1,3,1,1,2}
+98\,\HBS_{1,2,1,3,1,2,1}
+236\,\HBS_{1,2,1,3,2,1,1}
-176\,\HBS_{1,2,1,4,1,1,1}\nonumber\\[2mm]&&\hspace*{2mm}
-512\,\HBS_{1,2,2,1,1,1,3}
+604\,\HBS_{1,2,2,1,1,2,2}
-254\,\HBS_{1,2,2,1,1,3,1}
+156\,\HBS_{1,2,2,1,2,1,2}
+114\,\HBS_{1,2,2,1,2,2,1}\nonumber\\[2mm]&&\hspace*{2mm}
-220\,\HBS_{1,2,2,1,3,1,1}
+710\,\HBS_{1,2,2,2,1,1,2}
-364\,\HBS_{1,2,2,2,1,2,1}
-892\,\HBS_{1,2,2,2,2,1,1}
+386\,\HBS_{1,2,2,3,1,1,1}\nonumber\\[2mm]&&\hspace*{2mm}
-418\,\HBS_{1,2,3,1,1,1,2}
+210\,\HBS_{1,2,3,1,1,2,1}
+210\,\HBS_{1,2,3,1,2,1,1}
+462\,\HBS_{1,2,3,2,1,1,1}
-340\,\HBS_{1,2,4,1,1,1,1}\nonumber\\[2mm]&&\hspace*{2mm}
+480\,\HBS_{1,3,1,1,1,1,3}
-576\,\HBS_{1,3,1,1,1,2,2}
+250\,\HBS_{1,3,1,1,1,3,1}
-152\,\HBS_{1,3,1,1,2,1,2}
-194\,\HBS_{1,3,1,1,2,2,1}\nonumber\\[2mm]&&\hspace*{2mm}
+350\,\HBS_{1,3,1,1,3,1,1}
-152\,\HBS_{1,3,1,2,1,1,2}
+86\,\HBS_{1,3,1,2,1,2,1}
-254\,\HBS_{1,3,1,2,2,1,1}
+382\,\HBS_{1,3,1,3,1,1,1}\nonumber\\[2mm]&&\hspace*{2mm}
-382\,\HBS_{1,3,2,1,1,1,2}
+210\,\HBS_{1,3,2,1,1,2,1}
+210\,\HBS_{1,3,2,1,2,1,1}
+274\,\HBS_{1,3,2,2,1,1,1}
-116\,\HBS_{1,3,3,1,1,1,1}\nonumber\\[2mm]&&\hspace*{2mm}
+300\,\HBS_{1,4,1,1,1,1,2}
-156\,\HBS_{1,4,1,1,1,2,1}
-156\,\HBS_{1,4,1,1,2,1,1}
-156\,\HBS_{1,4,1,2,1,1,1}
-340\,\HBS_{1,4,2,1,1,1,1}\nonumber\\[2mm]&&\hspace*{2mm}
+240\,\HBS_{1,5,1,1,1,1,1}
+12\,\HBS_{2,1,1,1,1,2,3}
+12\,\HBS_{2,1,1,1,1,3,2}
-28\,\HBS_{2,1,1,1,1,4,1}
+18\,\HBS_{2,1,1,1,2,2,2}\nonumber\\[2mm]&&\hspace*{2mm}
+60\,\HBS_{2,1,1,1,2,3,1}
-162\,\HBS_{2,1,1,1,3,1,2}
+216\,\HBS_{2,1,1,1,3,2,1}
-124\,\HBS_{2,1,1,1,4,1,1}
-8\,\HBS_{2,1,1,2,1,2,2}\nonumber\\[2mm]&&\hspace*{2mm}
+14\,\HBS_{2,1,1,2,1,3,1}
+202\,\HBS_{2,1,1,2,2,1,2}
-390\,\HBS_{2,1,1,2,2,2,1}
+208\,\HBS_{2,1,1,2,3,1,1}
-188\,\HBS_{2,1,1,3,1,1,2}\nonumber\\[2mm]&&\hspace*{2mm}
+98\,\HBS_{2,1,1,3,1,2,1}
+236\,\HBS_{2,1,1,3,2,1,1}
-176\,\HBS_{2,1,1,4,1,1,1}
-8\,\HBS_{2,1,2,1,1,2,2}
+14\,\HBS_{2,1,2,1,1,3,1}\nonumber\\[2mm]&&\hspace*{2mm}
-64\,\HBS_{2,1,2,1,2,2,1}
+90\,\HBS_{2,1,2,1,3,1,1}
+198\,\HBS_{2,1,2,2,1,1,2}
-96\,\HBS_{2,1,2,2,1,2,1}
-466\,\HBS_{2,1,2,2,2,1,1}\nonumber\\[2mm]&&\hspace*{2mm}
+284\,\HBS_{2,1,2,3,1,1,1}
-176\,\HBS_{2,1,3,1,1,1,2}
+86\,\HBS_{2,1,3,1,1,2,1}
+86\,\HBS_{2,1,3,1,2,1,1}
+208\,\HBS_{2,1,3,2,1,1,1}\nonumber\\[2mm]&&\hspace*{2mm}
-156\,\HBS_{2,1,4,1,1,1,1}
-480\,\HBS_{2,2,1,1,1,1,3}
+552\,\HBS_{2,2,1,1,1,2,2}
-216\,\HBS_{2,2,1,1,1,3,1}
+152\,\HBS_{2,2,1,1,2,1,2}\nonumber\\[2mm]&&\hspace*{2mm}
+82\,\HBS_{2,2,1,1,2,2,1}
-196\,\HBS_{2,2,1,1,3,1,1}
+152\,\HBS_{2,2,1,2,1,1,2}
-86\,\HBS_{2,2,1,2,1,2,1}
+48\,\HBS_{2,2,1,2,2,1,1}\nonumber\\[2mm]&&\hspace*{2mm}
-180\,\HBS_{2,2,1,3,1,1,1}
+648\,\HBS_{2,2,2,1,1,1,2}
-350\,\HBS_{2,2,2,1,1,2,1}
-350\,\HBS_{2,2,2,1,2,1,1}
-770\,\HBS_{2,2,2,2,1,1,1}\nonumber\\[2mm]&&\hspace*{2mm}
+270\,\HBS_{2,2,3,1,1,1,1}
-384\,\HBS_{2,3,1,1,1,1,2}
+210\,\HBS_{2,3,1,1,1,2,1}
+210\,\HBS_{2,3,1,1,2,1,1}
+210\,\HBS_{2,3,1,2,1,1,1}\nonumber\\[2mm]&&\hspace*{2mm}
+470\,\HBS_{2,3,2,1,1,1,1}
-340\,\HBS_{2,4,1,1,1,1,1}
+480\,\HBS_{3,1,1,1,1,1,3}
-556\,\HBS_{3,1,1,1,1,2,2}
+222\,\HBS_{3,1,1,1,1,3,1}\nonumber\\[2mm]&&\hspace*{2mm}
-152\,\HBS_{3,1,1,1,2,1,2}
-162\,\HBS_{3,1,1,1,2,2,1}
+316\,\HBS_{3,1,1,1,3,1,1}
-152\,\HBS_{3,1,1,2,1,1,2}
+86\,\HBS_{3,1,1,2,1,2,1}\nonumber
\eeqa
\beqa
&&\hspace*{2mm}
-234\,\HBS_{3,1,1,2,2,1,1}
+376\,\HBS_{3,1,1,3,1,1,1}
-152\,\HBS_{3,1,2,1,1,1,2}
+86\,\HBS_{3,1,2,1,1,2,1}
+86\,\HBS_{3,1,2,1,2,1,1}\nonumber\\[2mm]&&\hspace*{2mm}
-274\,\HBS_{3,1,2,2,1,1,1}
+402\,\HBS_{3,1,3,1,1,1,1}
-384\,\HBS_{3,2,1,1,1,1,2}
+210\,\HBS_{3,2,1,1,1,2,1}
+210\,\HBS_{3,2,1,1,2,1,1}\nonumber\\[2mm]&&\hspace*{2mm}
+210\,\HBS_{3,2,1,2,1,1,1}
+270\,\HBS_{3,2,2,1,1,1,1}
-116\,\HBS_{3,3,1,1,1,1,1}
+300\,\HBS_{4,1,1,1,1,1,2}
-156\,\HBS_{4,1,1,1,1,2,1}\nonumber\\[2mm]&&\hspace*{2mm}
-156\,\HBS_{4,1,1,1,2,1,1}
-156\,\HBS_{4,1,1,2,1,1,1}
-156\,\HBS_{4,1,2,1,1,1,1}
-340\,\HBS_{4,2,1,1,1,1,1}
+240\,\HBS_{5,1,1,1,1,1,1}\nonumber\\[2mm]&&\hspace*{2mm}
-64\,\HBS_{1,1,1,1,1,2,2,2}
-40\,\HBS_{1,1,1,1,1,2,3,1}
+240\,\HBS_{1,1,1,1,1,3,1,2}
-296\,\HBS_{1,1,1,1,1,3,2,1}\nonumber\\[2mm]&&\hspace*{2mm}
+136\,\HBS_{1,1,1,1,1,4,1,1}
-344\,\HBS_{1,1,1,1,2,2,1,2}
+540\,\HBS_{1,1,1,1,2,2,2,1}
-220\,\HBS_{1,1,1,1,2,3,1,1}\nonumber\\[2mm]&&\hspace*{2mm}
+336\,\HBS_{1,1,1,1,3,1,1,2}
-164\,\HBS_{1,1,1,1,3,1,2,1}
-380\,\HBS_{1,1,1,1,3,2,1,1}
+256\,\HBS_{1,1,1,1,4,1,1,1}\nonumber\\[2mm]&&\hspace*{2mm}
+48\,\HBS_{1,1,1,2,1,2,2,1}
-76\,\HBS_{1,1,1,2,1,3,1,1}
-376\,\HBS_{1,1,1,2,2,1,1,2}
+172\,\HBS_{1,1,1,2,2,1,2,1}\nonumber\\[2mm]&&\hspace*{2mm}
+780\,\HBS_{1,1,1,2,2,2,1,1}
-496\,\HBS_{1,1,1,2,3,1,1,1}
+340\,\HBS_{1,1,1,3,1,1,1,2}
-156\,\HBS_{1,1,1,3,1,1,2,1}\nonumber\\[2mm]&&\hspace*{2mm}
-156\,\HBS_{1,1,1,3,1,2,1,1}
-360\,\HBS_{1,1,1,3,2,1,1,1}
+280\,\HBS_{1,1,1,4,1,1,1,1}
+48\,\HBS_{1,1,2,1,1,2,2,1}\nonumber\\[2mm]&&\hspace*{2mm}
-76\,\HBS_{1,1,2,1,1,3,1,1}
+204\,\HBS_{1,1,2,1,2,2,1,1}
-252\,\HBS_{1,1,2,1,3,1,1,1}
-336\,\HBS_{1,1,2,2,1,1,1,2}\nonumber\\[2mm]&&\hspace*{2mm}
+156\,\HBS_{1,1,2,2,1,1,2,1}
+156\,\HBS_{1,1,2,2,1,2,1,1}
+748\,\HBS_{1,1,2,2,2,1,1,1}
-380\,\HBS_{1,1,2,3,1,1,1,1}\nonumber\\[2mm]&&\hspace*{2mm}
+300\,\HBS_{1,1,3,1,1,1,1,2}
-156\,\HBS_{1,1,3,1,1,1,2,1}
-156\,\HBS_{1,1,3,1,1,2,1,1}
-156\,\HBS_{1,1,3,1,2,1,1,1}\nonumber\\[2mm]&&\hspace*{2mm}
-380\,\HBS_{1,1,3,2,1,1,1,1}
+280\,\HBS_{1,1,4,1,1,1,1,1}
+48\,\HBS_{1,2,1,1,1,2,2,1}
-76\,\HBS_{1,2,1,1,1,3,1,1}\nonumber\\[2mm]&&\hspace*{2mm}
+204\,\HBS_{1,2,1,1,2,2,1,1}
-252\,\HBS_{1,2,1,1,3,1,1,1}
+232\,\HBS_{1,2,1,2,2,1,1,1}
-156\,\HBS_{1,2,1,3,1,1,1,1}\nonumber\\[2mm]&&\hspace*{2mm}
-300\,\HBS_{1,2,2,1,1,1,1,2}
+156\,\HBS_{1,2,2,1,1,1,2,1}
+156\,\HBS_{1,2,2,1,1,2,1,1}
+156\,\HBS_{1,2,2,1,2,1,1,1}\nonumber\\[2mm]&&\hspace*{2mm}
+636\,\HBS_{1,2,2,2,1,1,1,1}
-380\,\HBS_{1,2,3,1,1,1,1,1}
+300\,\HBS_{1,3,1,1,1,1,1,2}
-156\,\HBS_{1,3,1,1,1,1,2,1}\nonumber\\[2mm]&&\hspace*{2mm}
-156\,\HBS_{1,3,1,1,1,2,1,1}
-156\,\HBS_{1,3,1,1,2,1,1,1}
-156\,\HBS_{1,3,1,2,1,1,1,1}
-380\,\HBS_{1,3,2,1,1,1,1,1}\nonumber\\[2mm]&&\hspace*{2mm}
+280\,\HBS_{1,4,1,1,1,1,1,1}
+48\,\HBS_{2,1,1,1,1,2,2,1}
-76\,\HBS_{2,1,1,1,1,3,1,1}
+204\,\HBS_{2,1,1,1,2,2,1,1}\nonumber\\[2mm]&&\hspace*{2mm}
-252\,\HBS_{2,1,1,1,3,1,1,1}
+232\,\HBS_{2,1,1,2,2,1,1,1}
-156\,\HBS_{2,1,1,3,1,1,1,1}
+156\,\HBS_{2,1,2,2,1,1,1,1}\nonumber\\[2mm]&&\hspace*{2mm}
-156\,\HBS_{2,1,3,1,1,1,1,1}
-300\,\HBS_{2,2,1,1,1,1,1,2}
+156\,\HBS_{2,2,1,1,1,1,2,1}
+156\,\HBS_{2,2,1,1,1,2,1,1}\nonumber\\[2mm]&&\hspace*{2mm}
+156\,\HBS_{2,2,1,1,2,1,1,1}
+156\,\HBS_{2,2,1,2,1,1,1,1}
+636\,\HBS_{2,2,2,1,1,1,1,1}
-380\,\HBS_{2,3,1,1,1,1,1,1}\nonumber\\[2mm]&&\hspace*{2mm}
+300\,\HBS_{3,1,1,1,1,1,1,2}
-156\,\HBS_{3,1,1,1,1,1,2,1}
-156\,\HBS_{3,1,1,1,1,2,1,1}
-156\,\HBS_{3,1,1,1,2,1,1,1}\nonumber\\[2mm]&&\hspace*{2mm}
-156\,\HBS_{3,1,1,2,1,1,1,1}
-156\,\HBS_{3,1,2,1,1,1,1,1}
-380\,\HBS_{3,2,1,1,1,1,1,1}
+280\,\HBS_{4,1,1,1,1,1,1,1}\nonumber\\[2mm]&&\hspace*{2mm}
-256\,\HBS_{1,1,1,1,1,2,2,1,1}
+392\,\HBS_{1,1,1,1,1,3,1,1,1}
-416\,\HBS_{1,1,1,1,2,2,1,1,1}
+280\,\HBS_{1,1,1,1,3,1,1,1,1}\nonumber\\[2mm]&&\hspace*{2mm}
-280\,\HBS_{1,1,1,2,2,1,1,1,1}
+280\,\HBS_{1,1,1,3,1,1,1,1,1}
-280\,\HBS_{1,1,2,2,1,1,1,1,1}
+280\,\HBS_{1,1,3,1,1,1,1,1,1}\nonumber\\[2mm]&&\hspace*{2mm}
-280\,\HBS_{1,2,2,1,1,1,1,1,1}
+280\,\HBS_{1,3,1,1,1,1,1,1,1}
-280\,\HBS_{2,2,1,1,1,1,1,1,1}
+280\,\HBS_{3,1,1,1,1,1,1,1,1}
\,.\label{ABArat}
\eeqa

\section{Analytical continuation of six-loop anomalous dimension at $\M=-2+\omega$} \label{AnConDL}

Analytic continuation of all harmonic sums up to transcendentality 11 was performed with the help of the \texttt{HARMPOL} package~\cite{Remiddi:1999ew} for \texttt{FORM}~\cite{Vermaseren:2000nd} and using \texttt{DATAMINE}~\cite{Blumlein:2009cf} tables for the substitution of the multiple zeta functions, or multiple polylogarithms at $x=1$ through usual Euler zeta-functions $\zeta_i$ and the minimal numbers of multiple zeta-functions. The results of the analytic continuation for the full planar six-loop anomalous dimension of twist-two operators in $\cN=4$ SYM theory is the following:
\beqa
\gamma(-2+\womega)&=&8 g^2\Big[
 -\frac{1}{\womega}
+ 1
+ \womega (1+{\z2})
+ \womega^2 (1-{\z3})
+ \womega^3 (1+{\z4})
+ \womega^4 (1-{\z5})\nonumber\\[.5mm]&&\hspace*{8mm}
+ \womega^5 (1+{\z6})
+ \womega^6 (1-{\z7})
+ \womega^7 (1+{\z8})
+ \womega^8 (1-{\z9})
\Big]\nonumber\\[.5mm]&&\hspace*{-14mm}
+g^4\bigg[
-\frac{32}{\womega^3}
+ \frac{64}{\womega^2}
+ \frac{32 {\z2}+32}{\womega}
+ -16 {\z3}
+ \womega (32 {\z2}-16 {\z3}-20 {\z4}-32)\nonumber\\[.5mm]&&\hspace*{-5mm}
+ \womega^2 (-64 {\z3} {\z2}+64 {\z2}-48 {\z3}+12 {\z4}+116 {\z5}-64)\nonumber\\[.5mm]&&\hspace*{-5mm}
+ \womega^3 \left(40 {\zp{3}{2}}-80 {\z3}+96 {\z2}+44 {\z4}-8 {\z5}-\frac{115 {\z6}}{3}-96\right)\nonumber\\[.5mm]&&\hspace*{-5mm}
+ \womega^4 \left(-128 {\z5} {\z2}+128 {\z2}-112 {\z3}-76 {\z3} {\z4}+76 {\z4}-40 {\z5}+5 {\z6}+\frac{579 {\z7}}{2}-128\right)\nonumber\\[.5mm]&&\hspace*{-5mm}
+ \womega^5 \Big(\frac{512 {\h}_{53}}{57}-\frac{2560 {\h}_{71}}{19}+160 {\z2}-144 {\z3}+108 {\z4}+\frac{3768 {\z3} {\z5}}{19}-72 {\z5}+37 {\z6}\nonumber\\[.5mm]&&\hspace*{10mm}
-3 {\z7}-\frac{16485 {\z8}}{76}-160\Big)\nonumber\\[.5mm]&&\hspace*{-5mm}
+ \womega^6 \Big(-192 {\z7} {\z2}+192 {\z2}-176 {\z3}+140 {\z4}-140 {\z4} {\z5}-104 {\z5}-69 {\z3} {\z6}+69 {\z6}\nonumber\\[.5mm]&&\hspace*{10mm}
-35 {\z7}+\frac{7 {\z8}}{4}+\frac{1057 {\z9}}{2}-192\Big)
\bigg]\nonumber\\[.5mm]&&\hspace*{-14mm}
+g^6\bigg[
-\frac{256}{\womega^5}
+ \frac{768}{\womega^4}
+ \frac{512 {\z2}}{\womega^3}
+ \frac{-768 {\z2}-384 {\z3}-512}{\womega^2}
+ \frac{-256 {\z2}+576 {\z3}-416 {\z4}-768}{\womega}\nonumber\\[.5mm]&&\hspace*{-5mm}
+ 448 {\z2} {\z3}+384 {\z3}+80 {\z4}+192 {\z5}-768
+ \womega \big(-336 \zp{3}{2}-704 {\z2} {\z3}+192 {\z3}-352 {\z4}\nonumber\\[.5mm]&&\hspace*{-5mm}
+272 {\z5}-1052 {\z6}-512\big)
+ \womega^2 \big(432 \zp{3}{2}-832 {\z2} {\z3}+104 {\z4} {\z3}-256 {\z2}-336 {\z4}\nonumber\\[.5mm]&&\hspace*{-5mm}
+1744 {\z2} {\z5}+1184 {\z5}+1616 {\z6}-1146 {\z7}\big)
+ \womega^3 \big(944 {\z2} \zp{3}{2}+656 \zp{3}{2}-1216 {\z2} {\z3}\nonumber\\[.5mm]&&\hspace*{-5mm}
-120 {\z4} {\z3}-\frac{44416 {\z5} {\z3}}{19}-192 {\z3}-\frac{8192 {\h_{53}}}{171}+\frac{40960 {\h_{71}}}{57}-768 {\z2}+128 {\z4}+1840 {\z5}\nonumber\\[.5mm]&&\hspace*{-5mm}
-1312 {\z2} {\z5}+\frac{268 {\z6}}{3}-584 {\z7}-\frac{12315 {\z8}}{19}+768\big)
+ \womega^4 \big(-432 \zp{3}{3}-384 {\z2} \zp{3}{2}+1008 \zp{3}{2}\nonumber\\[.5mm]&&\hspace*{-5mm}
-1856 {\z2} {\z3}-680 {\z4} {\z3}+\frac{50736 {\z5} {\z3}}{19}-\frac{6664 {\z6} {\z3}}{3}-384 {\z3}+\frac{10240 {\h_{53}}}{57}-\frac{51200 {\h_{71}}}{19}\nonumber\\[.5mm]&&\hspace*{-5mm}
-1536 {\z2}+1040 {\z4}-1376 {\z2} {\z5}-132 {\z4} {\z5}+2240 {\z5}-\frac{1816 {\z6}}{3}+5728 {\z2} {\z7}+2190 {\z7}\nonumber\\[.5mm]&&\hspace*{-5mm}
+\frac{90622 {\z8}}{171}-\frac{57904 {\z9}}{9}+1792\big)\bigg]\nonumber
\eeqa

\beqa
&&\hspace*{0mm}
+g^8\bigg[
-\frac{2560}{\womega^7}
+ \frac{10240}{\womega^6}
+ \frac{7168 {\z2}-5120}{\womega^5}
+ \frac{-18432 {\z2}-5632 {\z3}-10240}{\womega^4}\nonumber\\[1.5mm]&&\hspace*{6mm}
+ \frac{14336 {\z3}-13440 {\z4}-7680}{\womega^3}
+ \frac{9216 {\z3} {\z2}+8192 {\z2}+1536 {\z3}+14848 {\z4}+3200 {\z5}}{\womega^2}\nonumber\\[1.5mm]&&\hspace*{6mm}
+ \frac{-4416 \zp{3}{2}-14080 {\z2} {\z3}-6144 {\z3}+9216 {\z2}+3712 {\z4}-960 {\z5}-7504 {\z6}+10240}{\womega}\nonumber\\[1.5mm]&&\hspace*{6mm}
+ 6464 \zp{3}{2}-3328 {\z2} {\z3}-2400 {\z4} {\z3}-9728 {\z3}+6144 {\z2}+1024 {\z4}-1664 {\z2} {\z5}-4160 {\z5}\nonumber\\[1.5mm]&&\hspace*{6mm}
+\frac{45008 {\z6}}{3}+2472 {\z7}+20480
+ \womega \Big(1408 {\z2} \zp{3}{2}+1664 \zp{3}{2}+3072 {\z2} {\z3}-4512 {\z4} {\z3}\nonumber\\[1.5mm]&&\hspace*{6mm}
+\frac{10912 {\z5} {\z3}}{3}-10240 {\z3}+\frac{4096 {\h_{53}}}{27}-\frac{20480 {\h_{71}}}{9}+2048 {\z2}+2944 {\z4}+5120 {\z2} {\z5}\nonumber\\[1.5mm]&&\hspace*{6mm}
-6400 {\z5}+\frac{40640 {\z6}}{3}-8336 {\z7}+\frac{453500 {\z8}}{27}+28160\Big)
+ \womega^2 \Big(352 \zp{3}{3}+3456 {\z2} \zp{3}{2}\nonumber\\[1.5mm]&&\hspace*{6mm}
-1408 \zp{3}{2}+5120 {\z2} {\z3}-3776 {\z4} {\z3}-\frac{365344 {\z5} {\z3}}{57}-2520 {\z6} {\z3}-8704 {\z3}+5632 {\z4}\nonumber\\[1.5mm]&&\hspace*{6mm}
-\frac{372736 {\h_{53}}}{513}+\frac{1863680 {\h_{71}}}{171}+10880 {\z2} {\z5}+9496 {\z4} {\z5}-7680 {\z5}+5760 {\z6}\nonumber\\[1.5mm]&&\hspace*{6mm}
-38912 {\z2} {\z7}-29288 {\z7}-\frac{7149836 {\z8}}{513}+\frac{46024 {\z9}}{3}+30720\Big)
\bigg]\nonumber\\[1.5mm]&&\hspace*{0mm}
+g^{10}\bigg[
-\frac{28672}{\womega^9}
+ \frac{143360}{\womega^8}
+ \frac{102400 {\z2}-143360}{\womega^7}
+ \frac{-368640 {\z2}-81920 {\z3}-143360}{\womega^6}\nonumber\\[1.5mm]&&\hspace*{6mm}
+ \frac{163840 {\z2}+292864 {\z3}-289792 {\z4}}{\womega^5}\nonumber\\[1.5mm]&&\hspace*{6mm}
+ \frac{194560 {\z3} {\z2}+286720 {\z2}-110592 {\z3}+622080 {\z4}+40960 {\z5}+172032}{\womega^4}\nonumber\\[1.5mm]&&\hspace*{6mm}
+ \frac{286720-87040 \zp{3}{2}-473088 {\z2} {\z3}-241664 {\z3}+184320 {\z2}+37888 {\z4}-46080 {\z5}+\frac{137984}{3} {\z6}}{\womega^3}\nonumber\\[1.5mm]&&\hspace*{6mm}
+ \frac{209408 \zp{3}{2}+12288 {\z2} {\z3}-250368 {\z4} {\z3}-196608 {\z3}-150528 {\z4}-24064 {\z2} {\z5}-81920 {\z5}}{\womega^2}\nonumber\\[1.5mm]&&\hspace*{6mm}
+ \frac{\frac{767872}{3} {\z6}+36736 {\z7}+286720}{\womega^2}
+ \frac{126976 {\z2} \zp{3}{2}+10240 \zp{3}{2}+208896 {\z2} {\z3}+270592 {\z4} {\z3}}{\womega}\nonumber\\[1.5mm]&&\hspace*{6mm}
+ \frac{\frac{1537280}{57} {\z5} {\z3}-55296 {\z3}+\frac{843776}{513} {\h_{53}}-\frac{4218880}{171} {\h_{71}}-163840 {\z2}-144384 {\z4}-75264 {\z2} {\z5}}{\womega}\nonumber\\[1.5mm]&&\hspace*{6mm}
+ \frac{143360-71680 {\z5}-\frac{39424}{3} {\z6}-178752 {\z7}+\frac{144246544}{513} {\z8}}{\womega}
-41344 \zp{3}{3}-210432 {\z2} \zp{3}{2}-83456 \zp{3}{2}\nonumber\\[1.5mm]&&\hspace*{6mm}
+215040 {\z2} {\z3}+84480 {\z4} {\z3}+\frac{5259520}{57} {\z5} {\z3}-271040 {\z6} {\z3}+118784 {\z3}-245760 {\z2}-87552 {\z4} \nonumber\\[1.5mm]&&\hspace*{6mm}
-\frac{1163264}{513} {\h_{53}}+\frac{5816320}{171} {\h_{71}} +26624 {\z2} {\z5}-87968 {\z4} {\z5}-20480 {\z5}-\frac{657664}{3} {\z6}-83200 {\z2} {\z7}\nonumber\\[1.5mm]&&\hspace*{6mm}
+4032 {\z7}-\frac{101468128}{513} {\z8}-\frac{39344 {\z9}}{9}-143360
\bigg]\nonumber
\eeqa
\beqa
&&\hspace*{0mm}
+g^{12}\bigg[
-\frac{344064}{\womega^{11}}
+ \frac{2064384}{\womega^{10}}
+ \frac{1490944 {\z2}-3096576}{\womega^9}
+ \frac{-6881280 {\z2}-1204224 {\z3}-1376256}{\womega^8}\nonumber\\[1.5mm]&&\hspace*{6mm}
+ \frac{6307840 {\z2}+5529600 {\z3}-5625856 {\z4}+2064384}{\womega^7}\nonumber\\[1.5mm]&&\hspace*{6mm}
+ \frac{3768320 {\z3} {\z2}+5734400 {\z2}-4792320 {\z3}+17889280 {\z4}+546816 {\z5}+4128768}{\womega^6}\nonumber\\[1.5mm]&&\hspace*{6mm}
+ \frac{3440640-1640448 \zp{3}{2}-12771328 {\z2} {\z3}-5038080 {\z3}-6133760 {\z4}-1071104 {\z5}+3989504 {\z6}}{\womega^5}\nonumber\\[1.5mm]&&\hspace*{6mm}
+ \frac{5523456 \zp{3}{2}+5349376 {\z2} {\z3}-8453120 {\z4} {\z3}-737280 {\z3}-5505024 {\z2}-10680320 {\z4}}{\womega^4}\nonumber\\[1.5mm]&&\hspace*{6mm}
+ \frac{591360 {\z7}-667648 {\z2} {\z5}-1267712 {\z5}-2685952 {\z6}}{\womega^4}
+ \frac{3540992 {\z2} \zp{3}{2}-1994752 \zp{3}{2}}{\womega^3}\nonumber\\[1.5mm]&&\hspace*{6mm}
+ \frac{8953856 {\z2} {\z3}+17525760 {\z4} {\z3}+\frac{13147136}{19} {\z5} {\z3}+4276224 {\z3}+\frac{1638400}{57} {\h_{53}}-\frac{8192000}{19}{\h_{71}}}{\womega^3} \nonumber\\[1.5mm]&&\hspace*{6mm}
+ \frac{-8028160 {\z2}-7004160 {\z4}-849920 {\z2} {\z5}+100352 {\z5}-\frac{15717376}{3} {\z6}-3243520 {\z7}}{\womega^3}\nonumber\\[1.5mm]&&\hspace*{6mm}
+ \frac{\frac{745997696}{171} {\z8}-5160960}{\womega^3}
+ \frac{-1100800 \zp{3}{3}-8599552 {\z2} \zp{3}{2}-4012032 \zp{3}{2}+5283840 {\z2} {\z3}}{\womega^2}\nonumber\\[1.5mm]&&\hspace*{6mm}
+ \frac{495616 {\z4} {\z3}+\frac{47233024}{57} {\z5} {\z3}-\frac{5374720}{3} {\z6} {\z3}+7618560 {\z3}-\frac{37683200}{513} {\h_{53}}+\frac{188416000}{171}{\h_{71}}}{\womega^2} \nonumber\\[1.5mm]&&\hspace*{6mm}
+ \frac{-6881280 {\z2}-1880064 {\z4}+2869248 {\z2} {\z5}-1059328 {\z4} {\z5}+2027520 {\z5}-\frac{18255872}{3} {\z6}}{\womega^2}\nonumber\\[1.5mm]&&\hspace*{6mm}
+ \frac{-1978112 {\z2} {\z7}+1971200 {\z7}-\frac{6224639872}{513} {\z8}-\frac{607616}{9} {\z9}-10321920}{\womega^2}
\bigg]\,,
\eeqa
where
\beq
\h_{71}=H_{-7,-1}(1)
\,,\qquad
\h_{53}=H_{-5,-3}(1)\,,
\eeq
and $H_{i_1,...,i_k}(x)$ are the harmonic polylogarithms~\cite{Remiddi:1999ew}.

%

\begin{thebibliography}{10}

\bibitem{Maldacena:1997re}
J.~M. Maldacena, {\it {The large N limit of superconformal field theories and
  supergravity}},  {\em Adv. Theor. Math. Phys.} {\bf 2} (1998) 231--252,
  [\href{http://xxx.lanl.gov/abs/hep-th/9711200}{{\tt hep-th/9711200}}].

\bibitem{Gubser:1998bc}
S.~S. Gubser, I.~R. Klebanov, and A.~M. Polyakov, {\it {Gauge theory
  correlators from non-critical string theory}},  {\em Phys. Lett.} {\bf B428}
  (1998) 105--114, [\href{http://xxx.lanl.gov/abs/hep-th/9802109}{{\tt
  hep-th/9802109}}].

\bibitem{Witten:1998qj}
E.~Witten, {\it {Anti-de Sitter space and holography}},  {\em Adv. Theor. Math.
  Phys.} {\bf 2} (1998) 253--291,
  [\href{http://xxx.lanl.gov/abs/hep-th/9802150}{{\tt hep-th/9802150}}].

\bibitem{Berenstein:2002jq}
D.~E. Berenstein, J.~M. Maldacena, and H.~S. Nastase, {\it {Strings in flat
  space and pp waves from N=4 super Yang Mills}},  {\em JHEP} {\bf 04} (2002)
  013, [\href{http://xxx.lanl.gov/abs/hep-th/0202021}{{\tt hep-th/0202021}}].

\bibitem{Minahan:2002ve}
J.~A. Minahan and K.~Zarembo, {\it {The Bethe-ansatz for N=4 super
  Yang-Mills}},  {\em JHEP} {\bf 03} (2003) 013,
  [\href{http://xxx.lanl.gov/abs/hep-th/0212208}{{\tt hep-th/0212208}}].

\bibitem{Lipatov:1993yb}
L.~N. Lipatov, {\it {High-energy asymptotics of multicolor QCD and exactly
  solvable lattice models}},
  \href{http://xxx.lanl.gov/abs/hep-th/9311037}{{\tt hep-th/9311037}}.

\bibitem{Lipatov:1994xy}
L.~N. Lipatov, {\it {Asymptotic behavior of multicolor QCD at high energies in
  connection with exactly solvable spin models}},  {\em JETP Lett.} {\bf 59}
  (1994) 596--599. 

\bibitem{Faddeev:1994zg}
L.~D. Faddeev and G.~P. Korchemsky, {\it {High-energy QCD as a completely
  integrable model}},  {\em Phys. Lett.} {\bf B342} (1995) 311--322,
  [\href{http://xxx.lanl.gov/abs/hep-th/9404173}{{\tt hep-th/9404173}}].

\bibitem{Braun:1998id}
V.~M. Braun, S.~E. Derkachov, and A.~N. Manashov, {\it {Integrability of
  three-particle evolution equations in {QCD}}},  {\em Phys. Rev. Lett.} {\bf
  81} (1998) 2020--2023, [\href{http://xxx.lanl.gov/abs/hep-ph/9805225}{{\tt
  hep-ph/9805225}}].

\bibitem{Bena:2003wd}
I.~Bena, J.~Polchinski, and R.~Roiban, {\it {Hidden symmetries of the $AdS(5) \times
  S^5$ superstring}},  {\em Phys. Rev.} {\bf D69} (2004) 046002,
  [\href{http://xxx.lanl.gov/abs/hep-th/0305116}{{\tt hep-th/0305116}}].

\bibitem{Beisert:2003tq}
N.~Beisert, C.~Kristjansen, and M.~Staudacher, {\it {The dilatation operator of
  N=4 super Yang-Mills theory}},  {\em Nucl. Phys.} {\bf B664} (2003)
  131--184, [\href{http://xxx.lanl.gov/abs/hep-th/0303060}{{\tt
  hep-th/0303060}}].

\bibitem{Beisert:2003yb}
N.~Beisert and M.~Staudacher, {\it {The N=4 SYM integrable super spin chain}},
  {\em Nucl. Phys.} {\bf B670} (2003) 439--463,
  [\href{http://xxx.lanl.gov/abs/hep-th/0307042}{{\tt hep-th/0307042}}].

\bibitem{Beisert:2003jj}
N.~Beisert, {\it {The complete one-loop dilatation operator of N=4 Super
  Yang-Mills theory}},  {\em Nucl. Phys.} {\bf B676} (2004) 3--42,
  [\href{http://xxx.lanl.gov/abs/hep-th/0307015}{{\tt hep-th/0307015}}].

\bibitem{Beisert:2003ys}
N.~Beisert, {\it {The $su(2|3)$ dynamic spin chain}},  {\em Nucl. Phys.} {\bf
  B682} (2004) 487--520, [\href{http://xxx.lanl.gov/abs/hep-th/0310252}{{\tt
  hep-th/0310252}}].

\bibitem{Serban:2004jf}
D.~Serban and M.~Staudacher, {\it {Planar N=4 gauge theory and the Inozemtsev
  long range spin chain}},  {\em JHEP} {\bf 06} (2004) 001,
  [\href{http://xxx.lanl.gov/abs/hep-th/0401057}{{\tt hep-th/0401057}}].

\bibitem{Kazakov:2004qf}
V.~A. Kazakov, A.~Marshakov, J.~A. Minahan, and K.~Zarembo, {\it {
Classical/quantum integrability in AdS/CFT}},  {\em JHEP} {\bf 05} (2004) 024,
  [\href{http://xxx.lanl.gov/abs/hep-th/0402207}{{\tt hep-th/0402207}}].

\bibitem{Beisert:2004hm}
N.~Beisert, V.~Dippel, and M.~Staudacher, {\it {A novel long range spin chain
  and planar N=4 super Yang- Mills}},  {\em JHEP} {\bf 07} (2004) 075,
  [\href{http://xxx.lanl.gov/abs/hep-th/0405001}{{\tt hep-th/0405001}}].

\bibitem{Arutyunov:2004vx}
G.~Arutyunov, S.~Frolov, and M.~Staudacher, {\it {Bethe ansatz for quantum
  strings}},  {\em JHEP} {\bf 10} (2004) 016,
  [\href{http://xxx.lanl.gov/abs/hep-th/0406256}{{\tt hep-th/0406256}}].

\bibitem{Staudacher:2004tk}
M.~Staudacher, {\it {The factorized S-matrix of CFT/AdS}},  {\em JHEP} {\bf 05}
  (2005) 054, [\href{http://xxx.lanl.gov/abs/hep-th/0412188}{{\tt
  hep-th/0412188}}].

\bibitem{Beisert:2005fw}
N.~Beisert and M.~Staudacher, {\it {Long-range $PSU(2,2|4)$ Bethe ansaetze for
  gauge theory and strings}},  {\em Nucl. Phys.} {\bf B727} (2005) 1--62,
  [\href{http://xxx.lanl.gov/abs/hep-th/0504190}{{\tt hep-th/0504190}}].

\bibitem{Beisert:2005tm}
N.~Beisert, {\it {The $su(2|2)$ dynamic S-matrix}},  {\em Adv. Theor. Math.
  Phys.} {\bf 12} (2008) 945--979,
  [\href{http://xxx.lanl.gov/abs/hep-th/0511082}{{\tt hep-th/0511082}}].

\bibitem{Janik:2006dc}
R.~A. Janik, {\it {The $AdS(5) \times S^5$ superstring worldsheet S-matrix and
  crossing symmetry}},  {\em Phys. Rev.} {\bf D73} (2006) 086006,
  [\href{http://xxx.lanl.gov/abs/hep-th/0603038}{{\tt hep-th/0603038}}].

\bibitem{Arutyunov:2009kf}
  G.~Arutyunov and S.~Frolov,
  {\it {The dressing factor and crossing equations,}}
  {\em J.\ Phys.\ A} {\bf 42}  (2009) 425401,
  [\href{http://xxx.lanl.gov/abs/0904.4575}{{\tt arXiv:0904.4575}}].

\bibitem{Volin:2009uv}
  D.~Volin,
  {\it {Minimal solution of the AdS/CFT crossing equation}},
  {\em J.\ Phys.\ A} {\bf 42} (2009) 372001,
  [\href{http://xxx.lanl.gov/abs/0904.4929}{{\tt arXiv:0904.4929}}].

\bibitem{Hernandez:2006tk}
R.~Hernandez and E.~Lopez, {\it {Quantum corrections to the string Bethe
  ansatz}},  {\em JHEP} {\bf 07} (2006) 004,
  [\href{http://xxx.lanl.gov/abs/hep-th/0603204}{{\tt hep-th/0603204}}].

\bibitem{Arutyunov:2006iu}
G.~Arutyunov and S.~Frolov, {\it {On $AdS(5) \times S^5$ string S-matrix}},  {\em
  Phys. Lett.} {\bf B639} (2006) 378--382,
  [\href{http://xxx.lanl.gov/abs/hep-th/0604043}{{\tt hep-th/0604043}}].

\bibitem{Beisert:2006ib}
N.~Beisert, R.~Hernandez, and E.~Lopez, {\it {A crossing-symmetric phase for
  $AdS(5) \times S^5$ strings}},  {\em JHEP} {\bf 11} (2006) 070,
  [\href{http://xxx.lanl.gov/abs/hep-th/0609044}{{\tt hep-th/0609044}}].

\bibitem{Eden:2006rx}
B.~Eden and M.~Staudacher, {\it {Integrability and transcendentality}},  {\em
  J. Stat. Mech.} {\bf 0611} (2006) P11014,
  [\href{http://xxx.lanl.gov/abs/hep-th/0603157}{{\tt hep-th/0603157}}].

\bibitem{Bern:2006ew}
Z.~Bern, M.~Czakon, L.~J. Dixon, D.~A. Kosower, and V.~A. Smirnov, {\it {The
  four-loop planar amplitude and cusp anomalous dimension in maximally
  supersymmetric Yang-Mills theory}},  {\em Phys. Rev.} {\bf D75} (2007)
  085010, [\href{http://xxx.lanl.gov/abs/hep-th/0610248}{{\tt
  hep-th/0610248}}].

\bibitem{Beisert:2006ez}
N.~Beisert, B.~Eden, and M.~Staudacher, {\it {Transcendentality and crossing}},
   {\em J. Stat. Mech.} {\bf 0701} (2007) P01021,
  [\href{http://xxx.lanl.gov/abs/hep-th/0610251}{{\tt hep-th/0610251}}].

\bibitem{SchaferNameki:2006ey}
S.~Schafer-Nameki, M.~Zamaklar, and K.~Zarembo, {\it {How accurate is the
  quantum string Bethe ansatz?}},  {\em JHEP} {\bf 12} (2006) 020,
  [\href{http://xxx.lanl.gov/abs/hep-th/0610250}{{\tt hep-th/0610250}}].

\bibitem{Kotikov:2007cy}
A.~V. Kotikov, L.~N. Lipatov, A.~Rej, M.~Staudacher, and V.~N. Velizhanin, {\it
  {Dressing and wrapping}},  {\em J. Stat. Mech.} {\bf 0710} (2007) P10003,
  [\href{http://xxx.lanl.gov/abs/0704.3586}{{\tt arXiv:0704.3586}}].

\bibitem{Bajnok:2008bm}
Z.~Bajnok and R.~A. Janik, {\it {Four-loop perturbative Konishi from strings
  and finite size effects for multiparticle states}},  {\em Nucl. Phys.} {\bf
  B807} (2009) 625--650, [\href{http://xxx.lanl.gov/abs/0807.0399}{{\tt
  arXiv:0807.0399}}].

\bibitem{Bajnok:2008qj}
Z.~Bajnok, R.~A. Janik, and T.~Lukowski, {\it {Four loop twist two, BFKL,
  wrapping and strings}},  {\em Nucl. Phys.} {\bf B816} (2009) 376--398,
  [\href{http://xxx.lanl.gov/abs/0811.4448}{{\tt arXiv:0811.4448}}].

\bibitem{Bajnok:2009vm}
Z.~Bajnok, A.~Hegedus, R.~A. Janik, and T.~Lukowski, {\it {Five loop Konishi
  from AdS/CFT}},  {\em Nucl. Phys.} {\bf B827} (2010) 426--456,
  [\href{http://xxx.lanl.gov/abs/0906.4062}{{\tt arXiv:0906.4062}}].

\bibitem{Lukowski:2009ce}
T.~Lukowski, A.~Rej, and V.~N. Velizhanin, {\it {Five-loop anomalous dimension
  of twist-two operators}},  {\em Nucl. Phys.} {\bf B831} (2010) 105--132,
  [\href{http://xxx.lanl.gov/abs/0912.1624}{{\tt arXiv:0912.1624}}].

\bibitem{Gromov:2009tv}
N.~Gromov, V.~Kazakov, and P.~Vieira, {\it {Exact spectrum of anomalous
  dimensions of planar N=4 supersymmetric Yang-Mills theory}},  {\em Phys. Rev.
  Lett.} {\bf 103} (2009) 131601,
  [\href{http://xxx.lanl.gov/abs/0901.3753}{{\tt arXiv:0901.3753}}].

\bibitem{Gromov:2009bc}
N.~Gromov, V.~Kazakov, A.~Kozak, and P.~Vieira, {\it {Exact spectrum of
  anomalous dimensions of planar N = 4 supersymmetric Yang-Mills theory: TBA
  and excited states}},  {\em Lett. Math. Phys.} {\bf 91} (2010) 265--287,
  [\href{http://xxx.lanl.gov/abs/0902.4458}{{\tt arXiv:0902.4458}}].

\bibitem{Gromov:2009zb}
N.~Gromov, V.~Kazakov, and P.~Vieira, {\it {Exact spectrum of planar N=4 supersymmetric Yang- Mills theory: Konishi dimension at any coupling}},
   {\em Phys. Rev. Lett.} {\bf 104} (2010) 211601,
  [\href{http://xxx.lanl.gov/abs/0906.4240}{{\tt arXiv:0906.4240}}].

\bibitem{Bombardelli:2009ns}
D.~Bombardelli, D.~Fioravanti, and R.~Tateo, {\it {Thermodynamic Bethe Ansatz
  for planar AdS/CFT: a proposal}},  {\em J. Phys.} {\bf A42} (2009) 375401,
  [\href{http://xxx.lanl.gov/abs/0902.3930}{{\tt arXiv:0902.3930}}].

\bibitem{Arutyunov:2009ax}
G.~Arutyunov, S.~Frolov, and R.~Suzuki, {\it {Exploring the mirror TBA}},  {\em
  JHEP} {\bf 05} (2010) 031, [\href{http://xxx.lanl.gov/abs/0911.2224}{{\tt
  arXiv:0911.2224}}].

\bibitem{Arutyunov:2009ur}
G.~Arutyunov and S.~Frolov, {\it {Thermodynamic Bethe Ansatz for the $AdS_5
  \times S^5$ mirror model}},  {\em JHEP} {\bf 05} (2009) 068,
  [\href{http://xxx.lanl.gov/abs/0903.0141}{{\tt arXiv:0903.0141}}].

\bibitem{Arutyunov:2010gb}
G.~Arutyunov, S.~Frolov, and R.~Suzuki, {\it {Five-loop Konishi from the mirror
  TBA}},  {\em JHEP} {\bf 04} (2010) 069,
  [\href{http://xxx.lanl.gov/abs/1002.1711}{{\tt arXiv:1002.1711}}].

\bibitem{Balog:2010xa}
J.~Balog and A.~Hegedus, {\it {5-loop Konishi from linearized TBA and the XXX
  magnet}},  {\em JHEP} {\bf 06} (2010) 080,
  [\href{http://xxx.lanl.gov/abs/1002.4142}{{\tt arXiv:1002.4142}}].

\bibitem{Balog:2010vf}
J.~Balog and A.~Hegedus, {\it {The Bajnok-Janik formula and wrapping
  corrections}},  {\em JHEP} {\bf 09} (2010) 107,
  [\href{http://xxx.lanl.gov/abs/1003.4303}{{\tt arXiv:1003.4303}}].

\bibitem{Bajnok:2010ud}
Z.~Bajnok and O.~el~Deeb, {\it {6-loop anomalous dimension of a single impurity
  operator from AdS/CFT and multiple zeta values}},  {\em JHEP} {\bf 01} (2011)
  054, [\href{http://xxx.lanl.gov/abs/1010.5606}{{\tt arXiv:1010.5606}}].

\bibitem{Fiamberti:2007rj}
F.~Fiamberti, A.~Santambrogio, C.~Sieg, and D.~Zanon, {\it {Wrapping at four
  loops in N=4 SYM}},  {\em Phys. Lett.} {\bf B666} (2008) 100--105,
  [\href{http://xxx.lanl.gov/abs/0712.3522}{{\tt arXiv:0712.3522}}].

\bibitem{Fiamberti:2008sh}
F.~Fiamberti, A.~Santambrogio, C.~Sieg, and D.~Zanon, {\it {Anomalous dimension
  with wrapping at four loops in N=4 SYM}},  {\em Nucl. Phys.} {\bf B805}
  (2008) 231--266, [\href{http://xxx.lanl.gov/abs/0806.2095}{{\tt
  arXiv:0806.2095}}].

\bibitem{Velizhanin:2008jd}
V.~N. Velizhanin, {\it {The four-loop anomalous dimension of the Konishi
  operator in N=4 supersymmetric Yang-Mills theory}},  {\em JETP Lett.} {\bf
  89} (2009) 6--9, [\href{http://xxx.lanl.gov/abs/0808.3832}{{\tt
  arXiv:0808.3832}}].

\bibitem{Velizhanin:2008pc}
V.~N. Velizhanin, {\it {Leading transcedentality contributions to the four-loop
  universal anomalous dimension in N=4 SYM}},  {\em Phys. Lett.} {\bf B676}
  (2009) 112--115, [\href{http://xxx.lanl.gov/abs/0811.0607}{{\tt
  arXiv:0811.0607}}].

\bibitem{Lipatov:1976zz}
L.~N. Lipatov, {\it {Reggeization of the vector meson and the vacuum
  singularity in nonabelian gauge theories}},  {\em Sov. J. Nucl. Phys.} {\bf
  23} (1976) 338--345. 

\bibitem{Kuraev:1977fs}
E.~A. Kuraev, L.~N. Lipatov, and V.~S. Fadin, {\it {The Pomeranchuk singularity
  in nonabelian gauge theories}},  {\em Sov. Phys. JETP} {\bf 45} (1977)
  199--204. 

\bibitem{Balitsky:1978ic}
I.~I. Balitsky and L.~N. Lipatov, {\it {The Pomeranchuk singularity in Quantum
  Chromodynamics}},  {\em Sov. J. Nucl. Phys.} {\bf 28} (1978) 822--829.

\bibitem{Gorshkov:1966ht}
V.~G. Gorshkov, V.~N. Gribov, L.~N. Lipatov, and G.~V. Frolov, {\it {Doubly
  logarithmic asymptotic behavior in quantum electrodynamics}},  {\em Sov. J.
  Nucl. Phys.} {\bf 6} (1968) 95. 

\bibitem{Gorshkov:1966hu}
V.~G. Gorshkov, V.~N. Gribov, L.~N. Lipatov, and G.~V. Frolov, {\it {Backward
  electron - positron scattering at high- energies}},  {\em Sov. J. Nucl.
  Phys.} {\bf 6} (1968) 262. 

\bibitem{Gorshkov:1966qd}
V.~G. Gorshkov, V.~N. Gribov, L.~N. Lipatov, and G.~V. Frolov, {\it {Double
  logarithmic asymptotics of quantum electrodynamics}},  {\em Phys. Lett.} {\bf
  22} (1966) 671--673.

\bibitem{Kirschner:1982qf}
R.~Kirschner and L.~N. Lipatov, {\it {Double logarithmic asymptotics of quark
  scattering amplitudes with flavor exchange}},  {\em Phys. Rev.} {\bf D26}
  (1982) 1202--1205.

\bibitem{Kirschner:1982xw}
R.~Kirschner and L.~N. Lipatov, {\it {Doubly logarithmic asymptotic of the
  quark scattering amplitude with nonvacuum exchange in the t channel}},  {\em
  Sov. Phys. JETP} {\bf 56} (1982) 266--273.

\bibitem{Kirschner:1983di}
R.~Kirschner and L.~N. Lipatov, {\it {Double logarithmic asymptotics and Regge
  singularities of quark amplitudes with flavor exchange}},  {\em Nucl. Phys.}
  {\bf B213} (1983) 122--148.

\bibitem{Velizhanin:2011pb}
V.~N. Velizhanin, {\it {Double-logs, Gribov-Lipatov reciprocity and wrapping}},
   {\em JHEP} {\bf 08} (2011) 092,
  [\href{http://xxx.lanl.gov/abs/1104.4100}{{\tt arXiv:1104.4100}}].

\bibitem{Alfimov:2014bwa}
  M.~Alfimov, N.~Gromov and V.~Kazakov,
  {\it {QCD Pomeron from AdS/CFT Quantum Spectral Curve}}, 
  \href{http://arxiv.org/abs/1408.2530v3}{{\tt arXiv:1408.2530}}.

\bibitem{Gromov:2013pga}
N.~Gromov, V.~Kazakov, S.~Leurent, and D.~Volin, {\it {Quantum spectral curve
  for planar N=4 Super-Yang-Mills theory}},  {\em Phys.Rev.Lett.}
  {\bf 112} (2014), no.~1 011602,
  [\href{http://xxx.lanl.gov/abs/1305.1939}{{\tt arXiv:1305.1939}}].

\bibitem{Gromov:2014caa}
N.~Gromov, V.~Kazakov, S.~Leurent, and D.~Volin, {\it {Quantum spectral curve
  for arbitrary state/operator in $AdS_5/CFT_4$}},
  \href{http://xxx.lanl.gov/abs/1405.4857}{{\tt arXiv:1405.4857}}.

\bibitem{Velizhanin:2013vla}
V.~Velizhanin, {\it {Twist-2 at five loops: Wrapping corrections without
  wrapping computations}},  {\em JHEP} {\bf 1406} (2014) 108,
  [\href{http://xxx.lanl.gov/abs/1311.6953}{{\tt arXiv:1311.6953}}].

\bibitem{Marboe:2014gma}
C.~Marboe and D.~Volin, {\it {Quantum spectral curve as a tool for a
  perturbative quantum field theory}},
  \href{http://xxx.lanl.gov/abs/1411.4758}{{\tt arXiv:1411.4758}}.

\bibitem{Dippel}
V. Dippel, unpublished

\bibitem{Kotikov:2008pv}
A.~V. Kotikov, A.~Rej, and S.~Zieme, {\it {Analytic three-loop solutions for
  N=4 SYM twist operators}},  {\em Nucl. Phys.} {\bf B813} (2009) 460--483,
  [\href{http://xxx.lanl.gov/abs/0810.0691}{{\tt arXiv:0810.0691}}].

\bibitem{Kotikov:2002ab}
A.~V. Kotikov and L.~N. Lipatov, {\it {DGLAP and BFKL evolution equations in
  the N=4 supersymmetric gauge theory}},  {\em Nucl. Phys.} {\bf B661} (2003)
  19--61, [\href{http://xxx.lanl.gov/abs/hep-ph/0208220}{{\tt
  hep-ph/0208220}}]. 

\bibitem{Kotikov:2003fb}
A.~V. Kotikov, L.~N. Lipatov, and V.~N. Velizhanin, {\it {Anomalous dimensions
  of Wilson operators in N=4 SYM theory}},  {\em Phys. Lett.} {\bf B557}
  (2003) 114--120, [\href{http://xxx.lanl.gov/abs/hep-ph/0301021}{{\tt
  hep-ph/0301021}}].

\bibitem{Kotikov:2004er}
A.~V. Kotikov, L.~N. Lipatov, A.~I. Onishchenko, and V.~N. Velizhanin, {\it
  {Three-loop universal anomalous dimension of the Wilson operators in N=4
  SUSY Yang-Mills model}},  {\em Phys. Lett.} {\bf B595} (2004) 521--529,
  [\href{http://xxx.lanl.gov/abs/hep-th/0404092}{{\tt hep-th/0404092}}].
  [Erratum-ibid.B632:754-756,2006].

\bibitem{Moch:2004pa}
S.~Moch, J.~A.~M. Vermaseren, and A.~Vogt, {\it {The three-loop splitting
  functions in QCD: The non-singlet case}},  {\em Nucl. Phys.} {\bf B688}
  (2004) 101--134, [\href{http://xxx.lanl.gov/abs/hep-ph/0403192}{{\tt
  hep-ph/0403192}}].

\bibitem{Vogt:2004mw}
A.~Vogt, S.~Moch, and J.~A.~M. Vermaseren, {\it {The three-loop splitting
  functions in QCD: The singlet case}},  {\em Nucl. Phys.} {\bf B691} (2004)
  129--181, [\href{http://xxx.lanl.gov/abs/hep-ph/0404111}{{\tt
  hep-ph/0404111}}].

\bibitem{Vermaseren:1998uu}
J.~A.~M. Vermaseren, {\it {Harmonic sums, Mellin transforms and integrals}},
  {\em Int. J. Mod. Phys.} {\bf A14} (1999) 2037--2076,
  [\href{http://xxx.lanl.gov/abs/hep-ph/9806280}{{\tt hep-ph/9806280}}].

\bibitem{Dokshitzer:2005bf}
Y.~L. Dokshitzer, G.~Marchesini, and G.~P. Salam, {\it {Revisiting parton
  evolution and the large-x limit}},  {\em Phys. Lett.} {\bf B634} (2006)
  504--507, [\href{http://xxx.lanl.gov/abs/hep-ph/0511302}{{\tt
  hep-ph/0511302}}].

\bibitem{Dokshitzer:2006nm}
Y.~L. Dokshitzer and G.~Marchesini, {\it {N=4 SUSY Yang-Mills: Three loops
  made simple(r)}},  {\em Phys. Lett.} {\bf B646} (2007) 189--201,
  [\href{http://xxx.lanl.gov/abs/hep-th/0612248}{{\tt hep-th/0612248}}].

\bibitem{Basso:2006nk}
B.~Basso and G.~P. Korchemsky, {\it {Anomalous dimensions of high-spin
  operators beyond the leading order}},  {\em Nucl. Phys.} {\bf B775} (2007)
  1--30, [\href{http://xxx.lanl.gov/abs/hep-th/0612247}{{\tt hep-th/0612247}}].

\bibitem{Gribov:1972ri}
V.~N. Gribov and L.~N. Lipatov, {\it {Deep inelastic e p scattering in
  perturbation theory}},  {\em Sov. J. Nucl. Phys.} {\bf 15} (1972) 438--450.

\bibitem{Beccaria:2009eq}
M.~Beccaria, V.~Forini, T.~Lukowski, and S.~Zieme, {\it {Twist-three at five
  loops, Bethe Ansatz and wrapping}},  {\em JHEP} {\bf 03} (2009) 129,
  [\href{http://xxx.lanl.gov/abs/0901.4864}{{\tt arXiv:0901.4864}}].

\bibitem{Beccaria:2007bb}
M.~Beccaria, Y.~L. Dokshitzer, and G.~Marchesini, {\it {Twist 3 of the $sl(2)$
  sector of N=4 SYM and reciprocity respecting evolution}},  {\em Phys. Lett.}
  {\bf B652} (2007) 194--202, [\href{http://xxx.lanl.gov/abs/0705.2639}{{\tt
  arXiv:0705.2639}}].

\bibitem{Beccaria:2009vt}
M.~Beccaria and V.~Forini, {\it {Four loop reciprocity of twist two operators
  in N=4 SYM}},  {\em JHEP} {\bf 03} (2009) 111,
  [\href{http://xxx.lanl.gov/abs/0901.1256}{{\tt arXiv:0901.1256}}].

\bibitem{Lenstra:1982}
A.~Lenstra, H.~Lenstra, and L.~Lov{\'a}sz, {\it Factoring polynomials with
  rational coefficients},  {\em Math. Ann.} {\bf 261} (1982) 515--534.

\bibitem{Velizhanin:2010cm}
V.~N. Velizhanin, {\it {Six-loop anomalous dimension of twist-three operators
  in N=4 SYM}},  {\em JHEP} {\bf 11} (2010) 129,
  [\href{http://xxx.lanl.gov/abs/1003.4717}{{\tt arXiv:1003.4717}}].

\bibitem{Velizhanin:2012nm}
V.~N. Velizhanin, {\it {Three loop anomalous dimension of the non-singlet
  transversity operator in QCD}},  {\em Nucl. Phys.} {\bf B864} (2012)
  113--140, [\href{http://xxx.lanl.gov/abs/1203.1022}{{\tt arXiv:1203.1022}}].

\bibitem{MathematicaNotebook}
C.~Marboe and D.~Volin, {\it {{\texttt {MATHEMATICA}} notebook for the paper:
  "Quantum spectral curve as a tool for a perturbative quantum field theory"}},
 [\href{http://www.maths.tcd.ie/~dvolin/QSC/loop10sl2.zip}{{\tt http://www.maths.tcd.ie/~dvolin/QSC/loop10sl2.zip}}].

\bibitem{fplll}
M.~Albrech, D.~Cad\'{e}, X.~Pujol, and D.~Stehl\'{e}, {\it {fplll-4.0, a
  floating-point LLL implementation}},
  [\href{https://github.com/dstehle/fplll}{{\tt https://github.com/dstehle/fplll}}].

\bibitem{Fadin:1998py}
V.~S. Fadin and L.~N. Lipatov, {\it {BFKL pomeron in the next-to-leading
  approximation}},  {\em Phys. Lett.} {\bf B429} (1998) 127--134,
  [\href{http://xxx.lanl.gov/abs/hep-ph/9802290}{{\tt hep-ph/9802290}}].


\bibitem{Kotikov:2000pm}
A.~V. Kotikov and L.~N. Lipatov, {\it {NLO corrections to the BFKL equation in
  QCD and in supersymmetric gauge theories}},  {\em Nucl. Phys.} {\bf B582}
  (2000) 19--43, [\href{http://xxx.lanl.gov/abs/hep-ph/0004008}{{\tt
  hep-ph/0004008}}].

\bibitem{Remiddi:1999ew}
E.~Remiddi and J.~Vermaseren, {\it {Harmonic polylogarithms}},  {\em
  Int.J.Mod.Phys.} {\bf A15} (2000) 725--754,
  [\href{http://xxx.lanl.gov/abs/hep-ph/9905237}{{\tt hep-ph/9905237}}].

\bibitem{Vermaseren:2000nd}
J.~Vermaseren, {\it {New features of FORM}},
  \href{http://xxx.lanl.gov/abs/math-ph/0010025}{{\tt math-ph/0010025}}.

\bibitem{Blumlein:2009cf}
J.~Blumlein, D.~Broadhurst, and J.~Vermaseren, {\it {The multiple zeta value
  data mine}},  {\em Comput.Phys.Commun.} {\bf 181} (2010) 582--625,
  [\href{http://xxx.lanl.gov/abs/0907.2557}{{\tt arXiv:0907.2557}}].

\bibitem{Basso:2011rs}
  B.~Basso,
{\it{An exact slope for AdS/CFT}},
  [\href{http://xxx.lanl.gov/abs/1109.3154}{{\tt arXiv:1109.3154}}].
  
\bibitem{Basso:2012ex} 
  B.~Basso,
  {\it{Scaling dimensions at small spin in N=4 SYM theory,}}, 
  arXiv:1205.0054 [hep-th].

\bibitem{Gromov:2012eg} 
  N.~Gromov,
  {\it{``On the Derivation of the Exact Slope Function,}}, 
  JHEP {\bf 1302}, 055 (2013)
  [arXiv:1205.0018 [hep-th]].

\bibitem{Gromov:2014bva}
N.~Gromov, F.~Levkovich-Maslyuk, G.~Sizov and S.~Valatka,
{\it {Quantum spectral curve at work: from small spin to strong coupling in N=4 SYM}},
{\em JHEP} {\bf 1407} (2014) 156
  [\href{http://xxx.lanl.gov/abs/1402.0871}{{\tt arXiv:1402.0871}}].

\end{thebibliography}
%

\providecommand{\href}[2]{#2}

\begingroup\raggedright

\endgroup

\end{document}